\documentclass[12pt,preprint]{aastex}

\usepackage{amsmath}
\usepackage{amsfonts}
\usepackage{amssymb}
\usepackage{amsthm}
\usepackage{amscd}
\usepackage{eucal}
\usepackage{graphicx}
\usepackage{color}

\newcommand{\ubvri}{\protect\hbox{$U\!BV\!RI$} }
\newcommand{\bvri}{\protect\hbox{$BV\!RI$} }
\newcommand{\about}{$\sim$\!\!~}
\newcommand{\kms}{\,km\,s$^{-1}$}

\newcommand{\ssp}{\def\baselinestretch{1.0}\large\normalsize}

\shorttitle{SN~2002ap}
\shortauthors{Foley et~al.}

\begin{document}

\title{Optical Photometry and Spectroscopy of the SN~1998bw-like Type~Ic
Supernova 2002ap}

\vspace{2cm}

\author{Ryan J. Foley,\altaffilmark{1}
Marina S. Papenkova,\altaffilmark{1}
Brandon J. Swift,\altaffilmark{1}
Alexei V. Filippenko,\altaffilmark{1}
Weidong Li,\altaffilmark{1}
Paolo A. Mazzali,\altaffilmark{2}
Ryan Chornock,\altaffilmark{1}
Douglas C. Leonard,\altaffilmark{3} and
Schuyler D. Van Dyk\altaffilmark{4}}

\altaffiltext{1}{Department of Astronomy, 601 Campbell Hall,
University of California, Berkeley, CA 94720-3411;
\mbox{rfoley@astro.berkeley.edu}, \mbox{marina@ugastro.berkeley.edu},
\mbox{bswift@ugastro.berkeley.edu}, \mbox{alex@astro.berkeley.edu},
\mbox{weidong@astro.berkeley.edu}, \mbox{chornock@astro.berkeley.edu}}

\altaffiltext{2}{INAF-Osservatorio Astronomico, Via Giambattista
Tiepolo 11, 34131 Trieste, Italy; \mbox{mazzali@ts.astro.it}}

\altaffiltext{3}{Five College Astronomy Department, University of
Massachusetts, Amherst, MA 01003-9305; \mbox{leonard@nova.astro.umass.edu}}

\altaffiltext{4}{IPAC, California Institute of Technology, Mailcode
100-22, Pasadena, CA 91125; \mbox{vandyk@ipac.caltech.edu}}

\vspace{1cm}

\begin{abstract}
We present optical photometric and spectral data of the peculiar
Type~Ic SN~2002ap.  Photometric coverage includes \ubvri bands from
2002 January 30, the day after discovery, through 2002 December 12.
There are 5 early-time spectra and 8 in the nebular phase.  We
determine that SN~2002ap is similar to SN~1997ef and the
GRB-associated SN~1998bw with respect to spectral and photometric
characteristics.  The nebular spectra of SN~2002ap present the largest
\ion{Mg}{1}] $\lambda 4571$ to [\ion{O}{1}] $\lambda \lambda 6300$,
6364 ratio of any supernova spectra yet published, suggesting that the
progenitor of SN~2002ap was a highly stripped star.  Comparing the
nebular spectra of SN~1985F and SN~2002ap, we notice several similar
features, casting the classification of SN~1985F as a normal Type~Ib
supernova in doubt.  We also present nebular modeling of SN~2002ap and
find that the object ejected $\gtrsim \!\! 1.5~M_{\odot}$ of material
within the outer velocity shell of the nebula ($\sim 5500$ km
s$^{-1}$) and synthesized \about $0.09~M_{\odot}$ of $^{56}$Ni.
\end{abstract}

\medskip
\keywords{gamma-rays: bursts~--- line: identification~--- supernovae:
general~--- supernovae: individual (SNe~1985F, 1994I, 1998bw, 1999ex,
2002ap)}

\section{INTRODUCTION}\label{s:intro}

Core-collapse supernovae (SNe) are classified by their spectral (e.g.,
\citealt{Filippenko97}) and photometric properties (e.g.,
\citealt{Patat93}).  There are overwhelming indications that the
observational differences in these objects result from differences in
their progenitor stars.  Red supergiants with a full, massive hydrogen
envelope collapse to form SNe of Type~II plateau; the ``plateau''
refers to a period of relatively unchanged optical brightness after
maximum brightness.  SNe of Type~Ib occur through the same
core-collapse process as Type~II, except the progenitors are stripped
of their outer hydrogen layer by either stellar winds or transfer to a
companion; thus, the resulting spectra lack hydrogen.  If further
stripping occurs to the point of removing the deeper helium layer, the
SN is considered to be of Type Ic.

Recently, a subclass of SNe Ic has been identified as having unusually
high-velocity early-time ejecta reaching velocities of \about
30,000\kms \citep{Galama98}.  Models indicate that these objects can
have up to 10 times the kinetic energy per unit mass of normal
core-collapse SNe (\citealt{Iwamoto98}; see, however,
\citealt{Hoflich99} for a model in which this type of event is caused
by an aspherical explosion with normal SN~Ic energetics).  Besides
being among the most energetic events in the Universe, one particular
example of this subclass, SN~1998bw, has been associated with a
long-duration gamma-ray burst (GRB~980425; \citealt{Galama98}).  This
discovery strongly supports the hypothesis that the physical process
associated with long-duration GRBs (at least those of low gamma-ray
luminosity, like GRB~980425) is linked to the core collapse of
stripped stars.  Very recently, the optical afterglow of another, more
luminous GRB (GRB~030329) also exhibited weak spectral features
closely resembling those of SN~1998bw; accordingly, it has been given
a supernova designation (SN~2003dh; \citealt{Garnavich03a, Chornock03,
Stanek03}).

SN~2002ap was discovered on 2002 January 29.4 (UT dates will be used
throughout this paper) in the nearby spiral galaxy M74 by Y. Hirose
\citep{Nakano02}.  Reaching a peak brightness of $V \approx 12.4$ mag,
SN~2002ap was well observed.  Spectra were quickly obtained, and
SN~2002ap was identified as a SN Ic similar to SN~1998bw
\citep{Kinugasa02, Meikle02, Gal-Yam02b, Filippenko02}.

SN~2002ap has been closely followed in X-ray, radio, and optical
bands.  Specifically, \citet{Berger02} discovered synchrotron
self-absorption at radio wavelengths and \citet{Sutaria03} found
inverse-Compton scattering in X-rays.  At early times, SN~2002ap had
an inferred intrinsic continuum polarization of $\gtrsim \!\! 1\%$.
When modeled in terms of the oblate, electron-scattering atmospheres
of \citet{Hoflich91} this implies an asymmetry of at least 20\%
(\citealt{Leonard02}; see also \citealt{Kawabata02} and
\citealt{Wang03}).

\citet{Mazzali02} modeled the light curve and spectrum of SN~2002ap
using early-time data.  This model predicts a kinetic energy of \about
$(4 - 10) \times 10^{51}$ ergs, an ejected mass of $2.5-5$ $M_{\sun}$,
and a synthesized $^{56}$Ni mass of \about 0.07 $M_{\sun}$.  The model
also predicts that the progenitor of SN~2002ap had a main-sequence
mass of $20-25$ $M_{\sun}$. \citet{Maeda03b} fit a two-component model
to a bolometric light curve of SN~2002ap constructed from several
optical and infrared data sets.

In this paper, we present both early and late-time optical photometry
and spectroscopy of SN~2002ap. (Some preliminary results were
discussed by \citealt{Filippenko03}.) Along with an analysis of these
data and comparisons with other stripped core-collapse SNe, we
calculate a model of the nebular spectra, complementing the early-time
modeling of \citet{Mazzali02}.  In Section~\ref{s:photometry}, we
present photometry in \ubvri bands from the day after discovery and
lasting for \about 10 months.  In Section~\ref{s:spec}, we show 13
spectra ranging from 5 d to 386 d after $B$ maximum.
Section~\ref{s:model} contains the results of our nebular modeling.
We summarize our conclusions in Section~\ref{s:disc}.

\section{Photometry}\label{s:photometry}

\setcounter{footnote}{0}

\subsection{Photometric Observations and Data Reductions}

On 2002 January 30, one day after the discovery of SN~2002ap, the
Katzman Automatic Imaging Telescope \citep[KAIT;][]{Filippenko01} at
Lick Observatory began a \ubvri monitoring program that lasted until
2002 March 9 when solar conjunction prevented further observations.
After the SN became visible again in the morning skies, we obtained
additional \bvri photometry using the 1.0-m Nickel telescope at Lick
Observatory and the 1.5-m Oscar Meyer telescope at Palomar
Observatory.

KAIT obtained images of SN~2002ap (600~s in the $U$ band and 300~s in
\bvri bands) using an Apogee AP7 CCD camera, which has a
back-illuminated SITe $512 \times 512$ pixel CCD.  At the $f/8.2$
Cassegrain focus of KAIT, the 24~$\mu$m pixel of the chip yields a
scale of $0\arcsec.8$ pixel$^{-1}$, making the total field of view of
the camera $6\arcmin.7 \times 6\arcmin.7$.  The typical seeing at KAIT
is \about 3$\arcsec$ full width at half maximum (FWHM), so the CCD
images are well sampled.

For late-time observations, the camera at the Nickel telescope with a
back-illuminated $2048 \times 2048$ pixel Loral CCD was used in a
2$\times$2 binned mode, yielding a scale of $0\arcsec.36$ pixel$^{-1}$
and a total field of view of $6\arcmin.1\times 6\arcmin.1$
\citep{Li01}.  The typical images from the Nickel telescope have FWHM
$\approx 2\arcsec$, so once again the images are well sampled.

Late-time observations with the Palomar 1.5-m telescope were made with
the thinned $2048 \times 2048$ pixel ``CCD 13'' in the unbinned mode.
Each pixel is 24~$\mu$m in size for a scale of $0\arcsec.38$
pixel$^{-1}$.  Observations were made near an airmass of 1.06 with
exposure times 600~s and 300~s in Johnson $BV$ and Cousins $RI$
filters, respectively.  The typical seeing for these data was \about
$1.1\arcsec$, yielding well-sampled images.

For the KAIT images, dark-current subtraction and twilight-sky
flatfielding were done automatically at the telescope.  Since the KAIT
camera is thermoelectrically cooled, the camera temperature is
somewhat unstable, and thus the dark current is not always cleanly
subtracted from each image.  Consequently, there is a small
dark-current residual which varies from night to night that must be
removed manually by adding or subtracting fractions of a long-exposure
dark image.  Negligible uncertainties in the photometry are introduced
by the manual removal of the dark-current residuals since the SN and
comparison stars are rather bright \citep{Li01}.  Cosmic-ray removal
was accomplished with the IRAF\footnote{IRAF (Image Reduction and
Analysis Facility) is distributed by the National Optical Astronomy
Observatories, which are operated by the Association of Universities
for Research in Astronomy, Inc., under a cooperative agreement with
the National Science Foundation.} \emph{cosmicray} procedure in the
DAOPHOT package \citep{Stetson87}.  Reductions of the Lick Nickel and
Palomar observations were routine, not necessitating manual adjustment
of the dark-current image because the CCDs are cooled with liquid
nitrogen.


\begin{figure}
\ssp
\begin{center}
\scalebox{0.7}{
\plotone{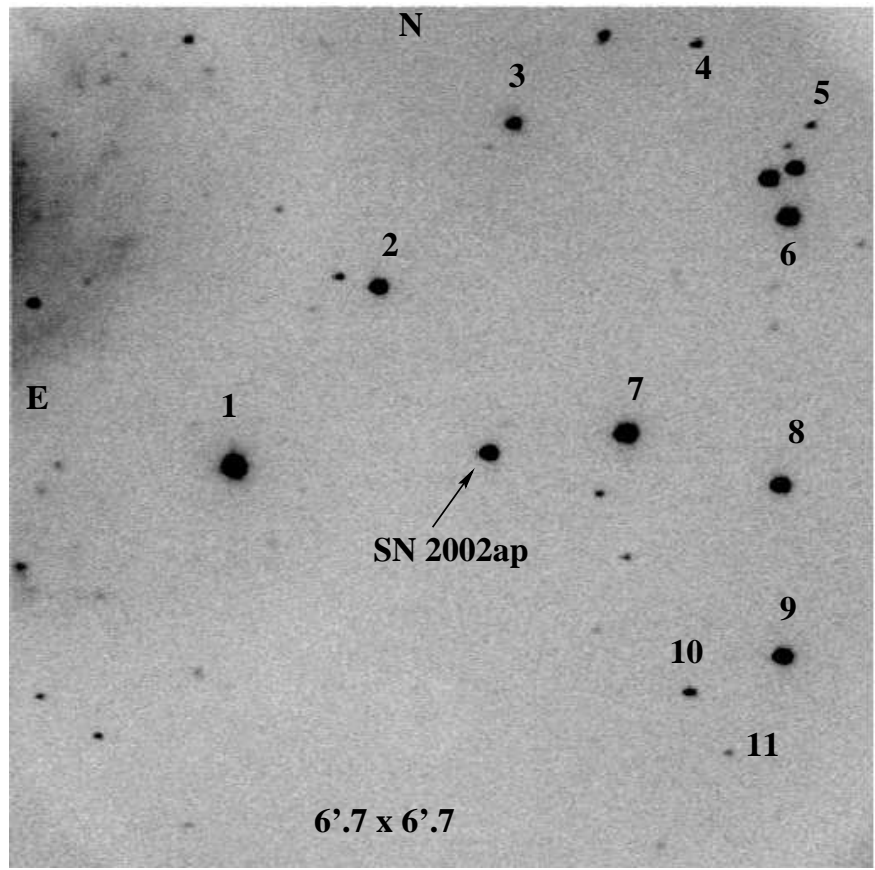}
}
\end{center}
\caption{$I$-band image of SN~2002ap and local standard stars, taken
on 2002 February 2.1.  A portion of the host galaxy M74 (NGC 628) is
barely visible in the northeast part of the field.}\label{f:field}
\end{figure}

Figure~\ref{f:field} shows a KAIT $I$-band image of SN~2002ap and 11
local standard stars taken on 2002 February 2.1.  Absolute calibration
of the field in \bvri was accomplished with the Nickel telescope on
2002 June 11, June 12, and July 9 under photometric
conditions. Calibrations for $U$ were obtained from \citet{Henden02}.
Table~\ref{t:locphot} shows photometry of the local standard stars
acquired from those observations.  \citet{Henden02} also provides
\bvri calibrations for the same stars which agree with our
calibrations very well (within \about 0.02 mag of each other).

\begin{deluxetable}{crrrrr}
\tabletypesize{\footnotesize}
\tablewidth{0pt}
\tablecaption{Photometry of Local Standard Stars\label{t:locphot}}
\tablehead{\colhead{Star ID} &
\colhead{$V$} &
\colhead{$U-B$} &
\colhead{$B-V$} &
\colhead{$V-R$} &
\colhead{$V-I$}}
\startdata

   1&12.955 (0.009)& 1.195 (0.032)&1.175 (0.025)&0.803 (0.020)&1.443 (0.006) \\
   2&13.824 (0.014)& 0.203 (0.023)&0.715 (0.013)&0.430 (0.010)&0.830 (0.012) \\
   3&14.624 (0.006)&$-$0.001 (0.015)&0.577 (0.008)&0.344 (0.005)&0.689 (0.012) \\
   4&17.221 (0.015)& 0.515 (0.047)&0.815 (0.084)&0.557 (0.047)&0.984 (0.037) \\
   5&18.044 (0.032)& 0.920 (0.151)&1.060 (0.084)&0.734 (0.030)&1.258 (0.035) \\
   6&13.248 (0.008)& 0.302 (0.020)&0.759 (0.009)&0.453 (0.009)&0.877 (0.006) \\
   7&13.111 (0.007)& 0.276 (0.013)&0.741 (0.007)&0.458 (0.009)&0.867 (0.001) \\
   8&13.713 (0.010)& 0.061 (0.016)&0.613 (0.006)&0.383 (0.005)&0.751 (0.006) \\
   9&13.944 (0.016)& 0.509 (0.014)&0.862 (0.008)&0.532 (0.019)&1.002 (0.016) \\
  10&16.851 (0.011)&$-$0.165 (0.157)&0.497 (0.033)&0.320 (0.027)&0.687 (0.028) \\
  11&18.492 (0.023)& 0.188 (0.087)&0.654 (0.069)&0.429 (0.100)&0.841 (0.068) \\

\enddata
\tablecomments{Units are magnitudes.  Uncertainties are indicated in
parentheses.}
\end{deluxetable}

Since SN~2002ap is located about $4.5\arcmin$ from the nucleus of M74
and far from any bright spiral arms, we used simple aperture
photometry to measure the instrumental magnitudes for SN~2002ap and
the local standard stars in the KAIT images, when the SN was still
bright. All local standard stars were measured except Stars 5 and 11,
which usually had poor signal to noise ratio in the KAIT images.  For
the $U$, $B$, and $V$ observations of 2002 January 30, Stars 2, 3, and
6 were not included in the images and thus were not measured.

For the Lick 1.0-m Nickel and Palomar 1.5-m observations, we applied
the point-spread-function (PSF) fitting technique to measure the
instrumental magnitudes of SN~2002ap and the local standard stars, as
the SN had already become quite faint in these images. The brightest
stars (e.g., Stars 1, 6, 7, 8, and 9) are often saturated in these
relatively deep images, and we used all the other unsaturated local
standard stars to construct a PSF for each image. During the
PSF-fitting process we also found that Stars 1 and 7 have nonstellar
PSFs when they are not saturated, especially in the $R$ and $I$ bands,
which are probably caused by the superimposition of some faint red
background stars.

The instrumental magnitudes thus measured are converted to the
standard Johnson-Cousins \ubvri system using the following
transformation equations:

\begin{align*}
u &= U + D_U(U-B) + C_U \\
b &= B + D_B(B-V) + C_B \\
v &= V + D_V(B-V) + C_V \\
r &= R + D_R(V-R) + C_R \\
i &= I + D_I(V-I) + C_I.
\end{align*}

Here, as in \citet{Modjaz00}, the lower-case bandpass letters signify
our instrumental magnitudes and the upper-case bandpass letters denote
magnitudes in the Johnson-Cousins system.  The constants $C_j$
indicate differences between the zero points of the instrumental and
standard systems, and the color-term coefficients $D_j$ for each
telescope and setup are summarized in Table~\ref{t:colors}.  Stars 1
and 7 were not used in the transformation due to their nonstellar
profile as noted above.

During the transformation process, we found some peculiarities for the
local standard Stars 2 and 3.  When differential photometry was
performed between SN~2002ap and these two stars for the KAIT images,
the resulting magnitudes for SN~2002ap were systematically different
(brighter by $> \!\!3\sigma$ and $> \!\!10\sigma$, respectively) from
these derived from the other comparison stars.  Yet, for the Lick
1.0-m and the Palomar 1.5-m images, both stars yielded measurements
consistent with the other stars.  We initially suspected that these
were two variable stars, but their brightness remained the same during
all the Nickel observations that spanned nearly 200 days.  Another
possible cause is that these two stars have peculiar colors, and when
observed with CCD cameras of different quantum efficiency curves,
their brightness may be systematically overestimated or
underestimated, but the two stars seem to have quite normal colors
compared to the other local standard stars.  We were thus unable to
find the cause of this inconsistency, and were forced to not use these
two stars in the transformation.  Fortunately, there are still enough
local standard stars in the field that provide consistent measurements
for the magnitudes of SN~2002ap, so we expect that our final
photometry is not affected by this inconsistency.

\begin{deluxetable}{cccccc}
\tabletypesize{\footnotesize}
\tablewidth{0pt}
\tablecaption{Summary of Color-Term Coefficients\label{t:colors}}
\tablehead{\colhead{Telescope} & \colhead{$D_U$} &\colhead{$D_B$} & \colhead{$D_V$} &
\colhead{$D_R$} & \colhead{$D_I$}}
\startdata
KAIT          &  $-$0.085 & $-$0.043 & 0.035 & 0.070 & $-$0.01  \\
Nickel 1.0-m  & \nodata & $-$0.080 & 0.060 & 0.100 & $-$0.035 \\
Palomar 1.5-m & \nodata & $-$0.080 & 0.015 & 0.120 & $-$0.115 \\
\enddata
\end{deluxetable}

Table~\ref{t:photometry} lists the final results from all of our
\ubvri observations of SN~2002ap, showing the early-time KAIT data and
the late-time Nickel and Palomar data.  Uncertainties, given in
parentheses, were determined by combining in quadrature the errors
given by the photometry routines in DAOPHOT with those introduced by
the transformation of instrumental magnitudes onto the standard
system. The uncertainties are dominated by the transformation errors
at all times except for the very latest Nickel observations when
SN~2002ap became fainter than 19 mag.

To double check our calibration of the SN~2002ap field and the final
photometry of SN~2002ap, we compared our results with those reported
in the literature. Our calibration of the SN~2002ap field is
consistent (to within $\pm$ 0.02 mag) with those reported by
\citet{Henden02}, \citet{Gal-Yam02a}, and \citet{Yoshii03}, including
Stars 2 and 3. Our $BVRI$ photometry of SN~2002ap is also generally
consistent with that reported by \citet{Yoshii03} to within $\pm$ 0.03
mag, although the difference in the $U$ band can be as large as 0.15
to 0.2 mag.  Notice that \citet{Yoshii03} used only one local standard
star (Star 7 in Figure~\ref{f:field}) which is probably contaminated
by other stars as discussed above, but they adopted an aperture that
is sufficiently large to include light in the whole system at all
times.

\subsection{Distance and Reddening}\label{ss:red}

\citet{Schlegel98} estimate the Galactic interstellar reddening in the
direction of M74 to be $E(B-V) = 0.071$ mag.  As shown in
Section~\ref{s:spec}, this value is in agreement with the color excess
derived from the narrow \ion{Na}{1} D lines in our spectra.  Reddening
of SN~2002ap from the gas in M74 is estimated by \citet{Klose02} to be
$E(B-V) = 0.008 \pm 0.002$ mag based on the \ion{Na}{1} D1 absorption
feature at the redshift of that galaxy in a spectrum of SN~2002ap.
The total extinction of $E(B-V) = 0.079$ mag is clearly dominated by
Galactic sources, which is to be expected in light of the projected
distance of SN~2002ap from the nucleus of M74.  The Galactic and host
extinction in each band were then calculated using conversion factors
in \citet{ODonnell94}, assuming a Galactic dust model for M74.  The
results of this procedure are summarized in Table~\ref{t:phottable}.

Currently, there is no Cepheid distance to M74.  Using photometry of
the brightest red and blue stars in the system, \citet{Sharina96} and
\citet{Sohn96} found distance moduli of 29.32 mag and 29.3 mag,
respectively.  This distance modulus corresponds to a distance of 7.3
Mpc.  Earlier studies of M74 give distance moduli which span a range
of about 5 mag, but the values from Sharina et~al.\ and
\citeauthor{Sohn96} are both in the middle of this range and very
consistent with each other.  The internal error given by Sharina
et~al.\ is 0.11 mag.  Sharina et~al.\ also determined that the mean
distance modulus for the M74 group is 29.46 mag differing by 0.14 mag
from their distance modulus for M74, resulting in a distance of 7.8
Mpc to the M74 group.  We will adopt the difference between the
distance modulus of M74 and its group (which is larger than the
internal error reported by Sharina et~al.) as the uncertainty in the
distance modulus.  The error in the distance modulus of \about 0.14
mag clearly overwhelms the uncertainty in the reddening (\about 0.01
mag) in any discussion of the absolute magnitude of SN~2002ap.  The
absolute magnitude of SN~2002ap derived from a distance modulus of
29.32 mag is presented in Table~\ref{t:phottable}.

\subsection{Optical Light Curves}

\begin{figure}
\ssp
\begin{center}
\rotatebox{90}{
\scalebox{0.7}{
\plotone{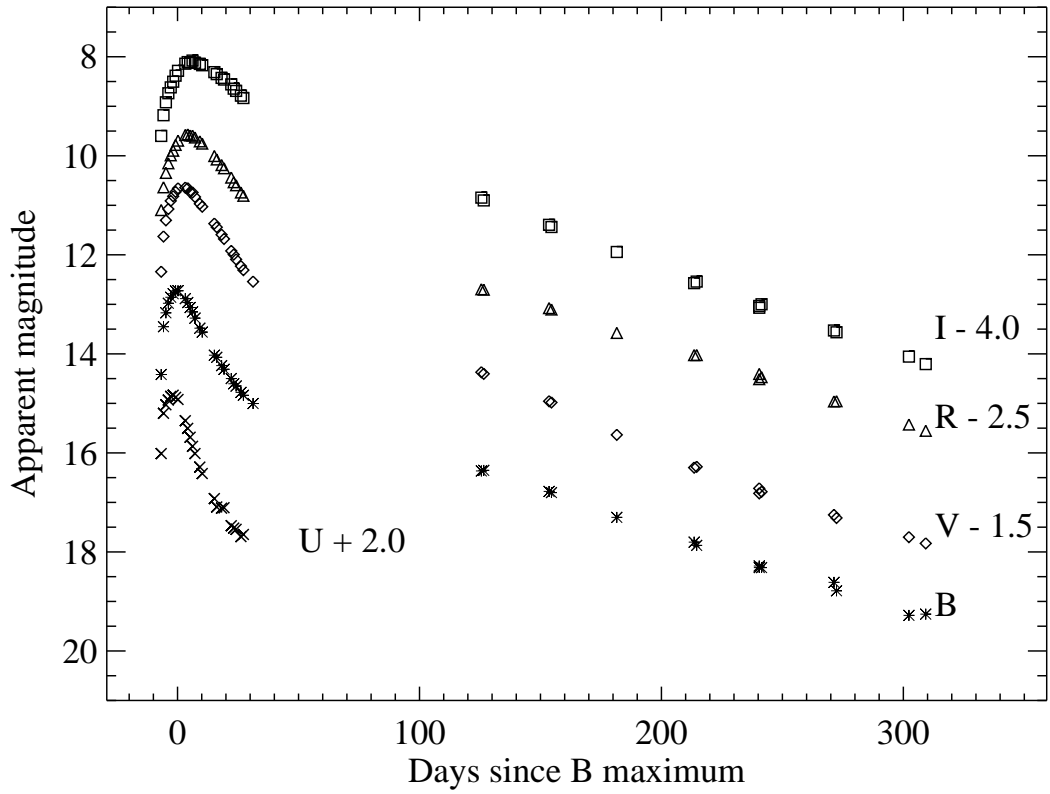}
}
}
\end{center}
\caption{The $U$, $B$, $V$, $R$, and $I$ light curves of SN~2002ap.
All measurements are from KAIT except for the points taken after solar
conjunction, which were obtained with the Lick Nickel 1-m and Palomar
1.5-m telescopes as indicated in Table~\ref{t:phottable}.  For the
majority of points, the uncertainties are smaller than the plotted
symbols.}\label{f:allcurves}
\end{figure}

Figure~\ref{f:allcurves} displays our \ubvri light curves of
SN~2002ap.  The maximum brightness and time of maximum brightness in
each band were estimated by fitting polynomials to the data points.
Since the light curves are well sampled near maximum, there is little
uncertainty in our results.  Table~\ref{t:phottable} lists the date
and magnitude of peak brightness in each band as determined by the
fits.  All relative dates are normalized such that $t = 0$ corresponds
to $B$~maximum on 2002 Feb $5.8 \pm 0.5$.  \citet{Mazzali02} indicate
that the explosion occurred on Jan $28.9 \pm 0.5$, \about 8 d before
$B$~maximum and \about 0.5 d before discovery.

\begin{deluxetable}{lccccc}
\tabletypesize{\scriptsize}
\tablewidth{0pt}
\tablecaption{Photometry of SN~2002ap\label{t:photometry}}
\tablehead{\colhead{JD -} & \colhead{$U$} &\colhead{$B$} & \colhead{$V$} &
\colhead{$R$} & \colhead{$I$}\\
\colhead{2452300} & \colhead{(mag)} & \colhead{(mag)} & \colhead{(mag)} & \colhead{(mag)} & \colhead{(mag)}}

\startdata

04.67&14.014 (0.074)&14.417 (0.023)&13.841 (0.026)&13.596 (0.024)&13.699 (0.015)\\
05.63&13.200 (0.031)&13.451 (0.015)&13.130 (0.021)&13.136 (0.020)&13.281 (0.018)\\
06.62&13.029 (0.031)&13.172 (0.018)&12.803 (0.021)&12.845 (0.024)&13.019 (0.029)\\
07.62&12.933 (0.037)&12.978 (0.021)&12.575 (0.020)&12.651 (0.017)&12.837 (0.024)\\
08.62&12.859 (0.039)&12.857 (0.017)&12.410 (0.016)&12.490 (0.016)&12.718 (0.023)\\
09.63&12.834 (0.033)&12.786 (0.019)&12.318 (0.028)&12.397 (0.021)&12.604 (0.017)\\
10.62&12.878 (0.039)&12.728 (0.019)&12.228 (0.024)&12.280 (0.015)&12.483 (0.023)\\
11.63&12.921 (0.039)&12.726 (0.028)&12.158 (0.017)&12.194 (0.018)&12.384 (0.026)\\
14.63&13.350 (0.034)&12.885 (0.013)&12.143 (0.018)&12.079 (0.023)&12.240 (0.019)\\
15.63&13.504 (0.045)&12.964 (0.018)&12.155 (0.020)&12.068 (0.019)&12.207 (0.020)\\
16.63&13.683 (0.050)&13.064 (0.015)&12.231 (0.021)&12.089 (0.016)&12.217 (0.018)\\
17.63&13.866 (0.058)&13.152 (0.023)&12.241 (0.024)&12.092 (0.016)&12.177 (0.027)\\
18.63&14.013 (0.041)&13.278 (0.016)&12.331 (0.020)&12.130 (0.018)&12.222 (0.026)\\
20.63&14.287 (0.041)&13.482 (0.017)&12.470 (0.022)&12.209 (0.018)&12.245 (0.023)\\
21.63&14.417 (0.047)&13.560 (0.026)&12.528 (0.026)&12.253 (0.024)&12.271 (0.022)\\
26.64&14.923 (0.057)&14.027 (0.021)&12.870 (0.024)&12.505 (0.021)&12.411 (0.022)\\
27.63&15.092 (0.040)&14.070 (0.024)&12.942 (0.023)&12.574 (0.023)&12.451 (0.020)\\
29.63&15.116 (0.051)&14.237 (0.029)&13.093 (0.023)&12.687 (0.019)&12.527 (0.024)\\
30.63&15.106 (0.066)&14.311 (0.023)&13.179 (0.025)&12.751 (0.018)&12.561 (0.027)\\
33.64&15.470 (0.037)&14.499 (0.040)&13.421 (0.026)&12.940 (0.018)&12.658 (0.025)\\
34.64&15.518 (0.059)&14.604 (0.023)&13.495 (0.022)&13.033 (0.022)&12.744 (0.022)\\
35.64&15.543 (0.066)&14.650 (0.026)&13.591 (0.027)&13.095 (0.031)&12.789 (0.026)\\
37.64&15.691 (0.075)&14.778 (0.023)&13.728 (0.029)&13.241 (0.023)&12.886 (0.015)\\
38.64&15.647 (0.087)&14.837 (0.023)&13.808 (0.033)&13.307 (0.024)&12.933 (0.020)\\
42.64&   \nodata    &15.003 (0.030)&14.041 (0.047)&    \nodata   &   \nodata\\
137.00\tablenotemark{a}&  \nodata  &16.364 (0.016)&15.870 (0.021)&15.195 (0.023)&14.945 (0.027)\\
137.99\tablenotemark{a}&  \nodata  &16.353 (0.020)&15.903 (0.017)&15.203 (0.020)&15.001 (0.017)\\
164.99\tablenotemark{a}&  \nodata  &16.778 (0.026)&16.457 (0.017)&15.576 (0.020)&15.494 (0.026)\\
165.99\tablenotemark{a}&  \nodata  &16.802 (0.024)&16.481 (0.015)&15.604 (0.025)&15.538 (0.029)\\
192.98\tablenotemark{a}&  \nodata  &17.302 (0.020)&17.135 (0.023)&16.077 (0.020)&16.042 (0.017)\\
224.96\tablenotemark{a}&  \nodata  &17.802 (0.023)&17.797 (0.017)&16.525 (0.018)&16.672 (0.019)\\
226.00\tablenotemark{a}&  \nodata  &17.868 (0.023)&17.781 (0.034)&16.524 (0.026)&16.640 (0.038)\\
251.85\tablenotemark{b}&  \nodata  &18.284 (0.032)&18.217 (0.027)&17.011 (0.030)&17.133 (0.064)\\
251.87\tablenotemark{b}&  \nodata  &18.316 (0.032)&18.317 (0.027)&16.906 (0.030)&17.169 (0.064)\\
252.82\tablenotemark{b}&  \nodata  &18.316 (0.029)&18.285 (0.041)&16.970 (0.037)&17.098 (0.064)\\
282.75\tablenotemark{a}&  \nodata  &18.614 (0.096)&18.751 (0.043)&17.460 (0.038)&17.628 (0.053)\\
283.80\tablenotemark{a}&  \nodata  &18.785 (0.073)&18.814 (0.037)&17.456 (0.037)&17.664 (0.047)\\
313.74\tablenotemark{a}&  \nodata  &19.281 (0.095)&19.201 (0.050)&17.927 (0.062)&18.156 (0.063)\\
320.69\tablenotemark{a}&  \nodata  &19.256 (0.091)&19.328 (0.083)&18.053 (0.050)&18.307 (0.071)\\

\enddata
\tablecomments{Uncertainties are indicated in parentheses. Corrected
for the reddening indicated in Table~\ref{t:phottable}.}
\tablenotetext{a}{Lick Observatory, Nickel 1-m telescope.}
\tablenotetext{b}{Palomar Observatory, Oscar Meyer 1.5-m telescope.}  
\end{deluxetable}

\begin{deluxetable}{cccccc}
\tabletypesize{\footnotesize}
\tablewidth{0pt}
\tablecaption{SN~2002ap Maximum Brightness, Corrected for Reddening\label{t:phottable}}
\tablehead{\colhead{Filter} & \colhead{$U$} & \colhead{$B$} & \colhead{$V$} &
\colhead{$R$} & \colhead{$I$}}

\startdata
UT Date (2002) & Feb 4.1 & Feb 5.8 & Feb 7.8 & Feb 10.1 & Feb 11.5\\
JD -- 2452300  & 9.61 (0.08) & 11.43 (0.22) & 13.42 (0.16) & 15.74 (0.18)
& 17.02 (0.20) \\
Days past Explosion\tablenotemark{a}& 6.2 & 7.9 & 9.9 & 12.2 & 13.6 \\
App. Mag & 13.245 (0.005) & 13.059 (0.005) & 12.387 (0.006) & 12.267
(0.003) & 12.323 (0.005) \\
Adopted Extinction & 0.413 & 0.348 & 0.263 & 0.199 & 0.127 \\
Abs. Mag & $-16.1$ (0.2) & $-16.3$ (0.2) & $-16.9$ (0.2) & $-17.1$ (0.2) & $-17.0$ (0.2) \\
\enddata
\tablecomments{Uncertainties are indicated in parentheses.}
\tablenotetext{a}{Date of explosion is 2002 Jan. 28.9 (JD =2452303.4)
as calculated by \citet{Mazzali02}.}
\end{deluxetable}


In the following sections, we discuss the behavior of SN~2002ap in
each bandpass relative to the stripped core-collapse SNe~1994I
\citep{Richmond96, Richmond03}, 1999ex \citep{Stritzinger02}, 1998bw
\citep{Galama98, Sollerman02}, 1997ef \citep{Garnavich03b}, and 1985F
\citep{Tsvetkov86}, when the data are available in that bandpass.  In
each of the comparison plots (Figures~\ref{f:uband} through
\ref{f:iband}) the light curves have been normalized to the time of
$B$ maximum for each supernova and to the same peak magnitude in the
plotted band.

Since SN~2002ap exhibits unusually high expansion velocities in its
spectra at early times similar to those of the GRB-associated
SN~1998bw, we include the light curves of SN~1998bw for comparison in
\bvri\!\!\!. We also compare SN~2002ap to the prototypical SN~Ic~1994I
and the SN~Ib/c~1999ex in \ubvri\!\!\!.  Because of limited available
data, we display only the $V$-band light curve of the high-velocity
SN~Ic~1997ef, another peculiarly energetic event like SN~1998bw.
Because of a lack of data in other bands, we only examine the $B$-band
light curve of SN~1985F, which has a similar nebular-phase spectrum to
that of SN~2002ap.

\subsubsection{$U$~Band}\label{ss:uband}

 
\begin{figure}
\ssp
\begin{center}
\rotatebox{90}{
\scalebox{0.7}{
\plotone{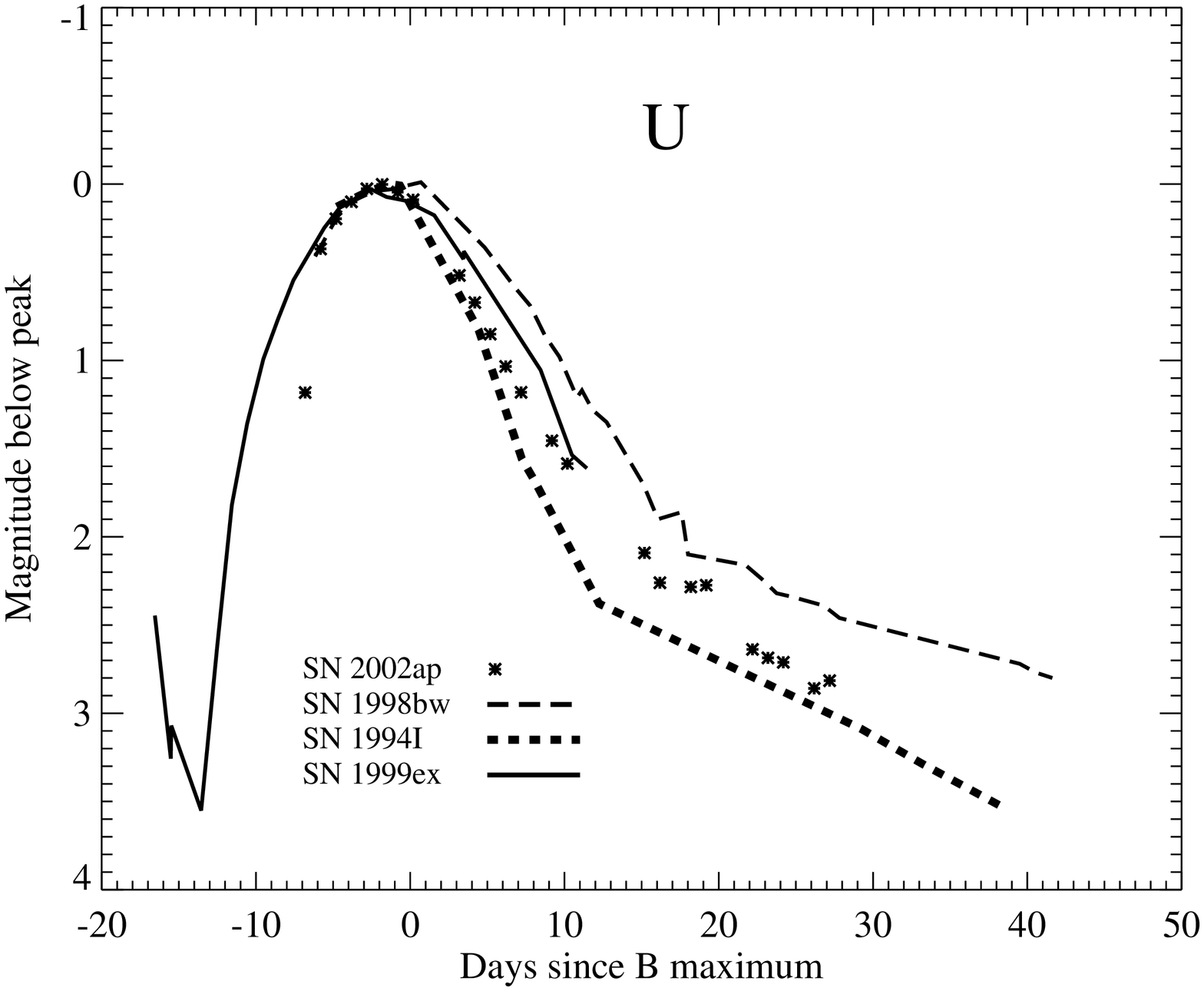}
}
}
\end{center}
\caption{The $U$-band light curve of SN~2002ap together with those of
SNe~1994I \citep{Richmond03}, 1999ex \citep{Stritzinger02}, and 1998bw
\citep{Galama98}.  All of the curves are shifted to match the time of
$B$ maximum and the same peak magnitude in the $U$
band.}\label{f:uband}
\end{figure}

Figure~\ref{f:uband} shows the $U$-band light curve of SN~2002ap with
the light curves of SNe~1994I, 1999ex, and 1998bw.  Before maximum at
$t = -1.8$ d, SN~2002ap rose more quickly than SN~1999ex.  After
maximum brightness, SN~1998bw declined slower than any of these, while
SN~2002ap declined slightly faster than SN~1999ex until $t = 11$ d
when observations of SN~1999ex ceased.  SN~1994I declined more quickly
than all other SNe.

\subsubsection{$B$~Band}\label{ss:bband}


\begin{figure}
\ssp
\begin{center}
\rotatebox{90}{
\scalebox{0.7}{
\plotone{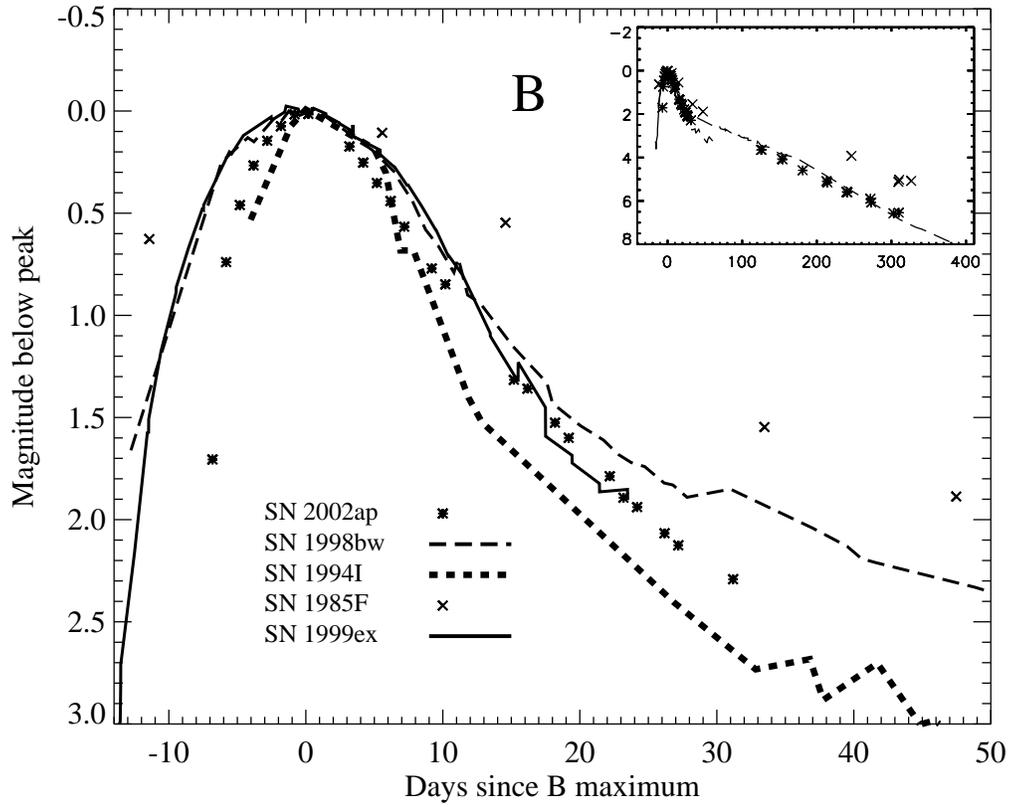}
}
}
\end{center}
\caption{The $B$-band light curve of SN~2002ap, together with those of
SNe~1994I \citep{Richmond96}, 1999ex \citep{Stritzinger02}, 1998bw
\citep{Galama98, Sollerman02}, and 1985F \citep{Tsvetkov86}.  All of
the curves are shifted to match the time of $B$ maximum and the same
peak magnitude in this band.  The inset shows the late-time
decline.}\label{f:bband}
\end{figure}

Figure~\ref{f:bband} shows the $B$-band light curve of SN~2002ap with
the light curves of SNe~1994I, 1999ex, 1998bw, and 1985F.  Before
maximum, SN~2002ap rises faster than SN~Ib/c~1999ex and the
high-velocity SN~Ic~1998bw.  It appears that SN~2002ap rises at the
same rate as (or slower than) SN~Ic~1994I, but this conclusion is not
definitive because of a relatively late discovery of SN~1994I.

>From maximum brightness until $t \approx 7$ d, SN~2002ap declines
similarly to SN~1994I, and SNe~1998bw and 1999ex decline more slowly.
After $t \approx 9$ d, the decline rate of SN~2002ap is relatively
stable, while SN~1994I starts declining more rapidly.  SN~2002ap
follows SN~1998bw until around $t = 20$ d (at a slightly fainter
magnitude relative to the peak) when SN~1998bw begins its slower
late-time decline phase.  SN~1999ex closely follows SN~1998bw until $t
= 12$ d, after which it begins fading more rapidly to follow the light
curve of SN~2002ap.

Notice that SN~1985F has an extremely wide $B$ light curve relative to
even the energetic SNe~1998bw and 2002ap.  The slow rise and decline
of the light curve suggest high energetics and a large $^{56}$Ni mass
as in SN~1997ef (\citealt{Iwamoto00}; cf. the $V$ light curve of
SN~1997ef in Section~\ref{ss:vband}).

At late times, SN~2002ap declines at a mean rate of $0.017 \pm 0.001$
and $0.016 \pm 0.001$ mag day$^{-1}$ at $120 < t \lesssim 220$ d and
$220 \lesssim t < 310$ d, respectively.  Data are only available at
late times for SN~1998bw in $B$ until $t \approx 190$ d ($0.014 \pm
0.001$ mag day$^{-1}$) and after $t \approx 320$ d ($0.015 \pm 0.001$
mag day$^{-1}$).  SN~2002ap seems to have faded more quickly than
SN~1998bw at late times.  We cannot conclude that SN~2002ap faded more
quickly than SN~1998bw during the interval $200-300$ d because of a
lack of data.  In addition, SN~1985F declines at a rate of about
$0.012 \pm 0.002$ mag day$^{-1}$ at late times, much slower than
SNe~1998bw and 2002ap, and at $t \approx 300$ d is still about 1.5 mag
brighter than either of these SNe.  However, the early data for
SN~1985F are sparse, which may have resulted in a poor calculation of
its peak magnitude.

\subsubsection{$V$~Band}\label{ss:vband}


\begin{figure}
\ssp
\begin{center}
\rotatebox{90}{
\scalebox{0.7}{
\plotone{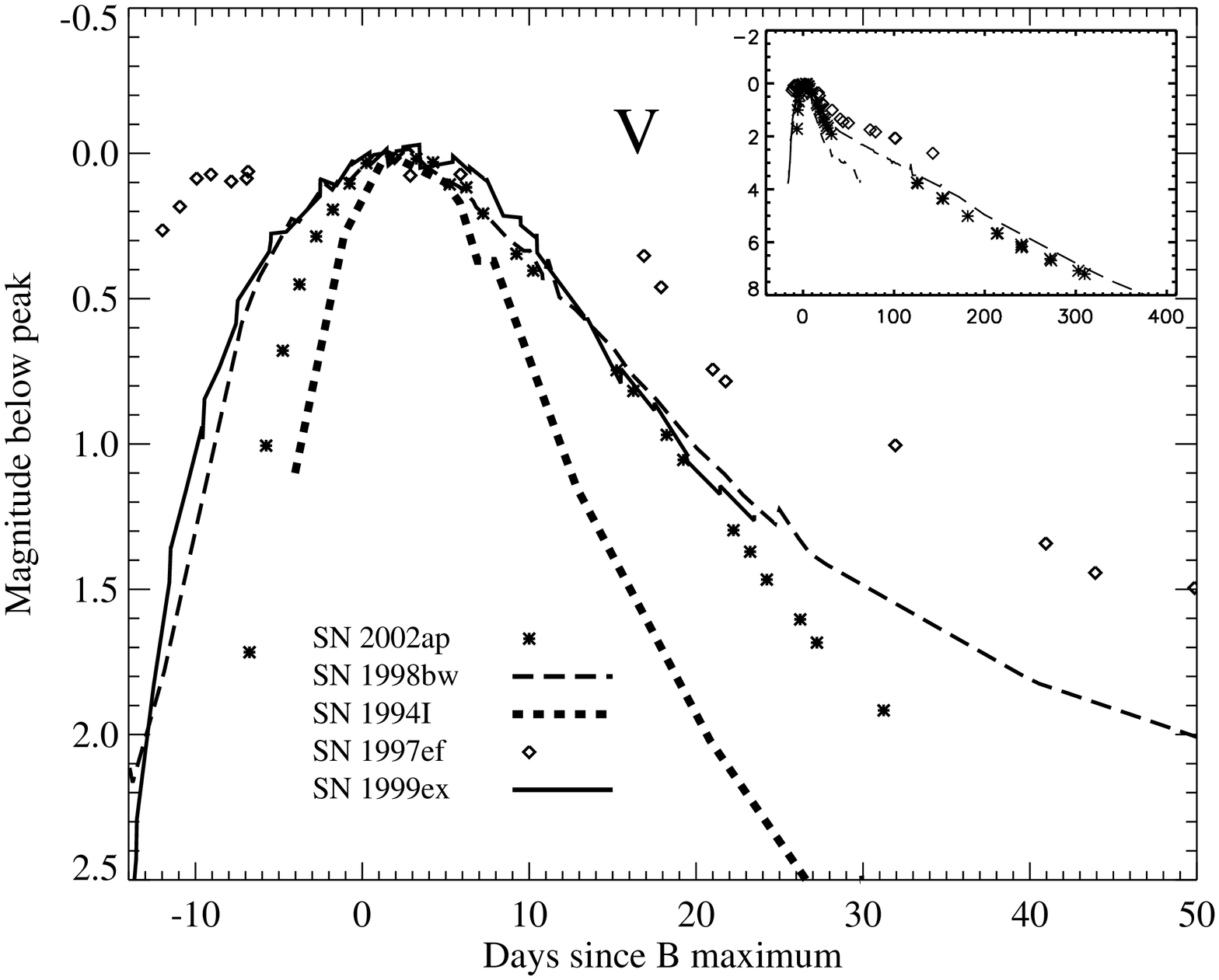}
}
}
\end{center}
\caption{Same as Figure~\ref{f:bband}, but for the $V$-band light
curve of SN~2002ap.  We also include that of SN~1997ef
\citep{Garnavich03b} and exclude SN~1985F.}\label{f:vband}
\end{figure}

We present the SN~2002ap $V$-band light curve along with other
comparison curves in Figure~\ref{f:vband}.  Before $V$ maximum at $t
\approx -3$ d, SNe~2002ap and 1994I rise much more quickly than
SNe~1999ex and 1999bw; as in $B$, however, SN~2002ap rises slightly
more slowly than SN~1994I.

After maximum brightness ($t = 3-20$ d), SNe~2002ap, 1999ex, and
1998bw decline in a similar manner.  The decline occurs at a
significantly slower rate than the rise, unlike for SN~1994I, which
fades precipitously after maximum.  At $t > 20$ d, SN~1998bw fades at
a slower rate as it enters its late-time decline phase; meanwhile
SN~2002ap continues its rapid decline.

SNe~2002ap, 1998bw, and 1997ef also have similar peaks, being
noticeably broader than that of SN~ 1994I.  Such broad peaks seem to
be a characteristic of the high-velocity SN~Ic ``hypernovae,'' as
noted for SN~1997ef by \citet{Iwamoto00}.  However, as discussed by
\citet{Clocchiatti97}, SN~Ib/c light curves are heterogeneous,
consisting of both fast and slow decliners.  \citet{Clocchiatti97}
determined SN~1994I to be a fast decliner, while \citet{Stritzinger02}
established that SN~1999ex is a slow decliner.  Most SNe~Ic tend to be
fast decliners, but SN~1990B is an example of a slow-declining SN~Ic
\citep{Clocchiatti01}.  This heterogeneity is apparent from the light
curves presented here.

During times $120 - 220$ d and $220 - 310$ d, SN~2002ap fades at a
mean rate of $0.022 \pm 0.001$ and $0.017 \pm 0.001$ mag day$^{-1}$
(respectively) in the $V$ band.  SN~1998bw, on the other hand, fades
at a nearly constant rate of $0.019 \pm 0.001$ mag day$^{-1}$ during
the interval $50 - 380$ d, so SNe~2002ap and 1998bw decline at nearly
the same rate here during the nebular phase.  In the early nebular
phase, $70 < t < 140$ d, SN~1997ef declines at a mean rate of $0.012
\pm 0.001$ mag day$^{-1}$, much slower than either of the other two
SNe.

\subsubsection{$R$~Band}


\begin{figure}
\ssp
\begin{center}
\rotatebox{90}{
\scalebox{0.7}{
\plotone{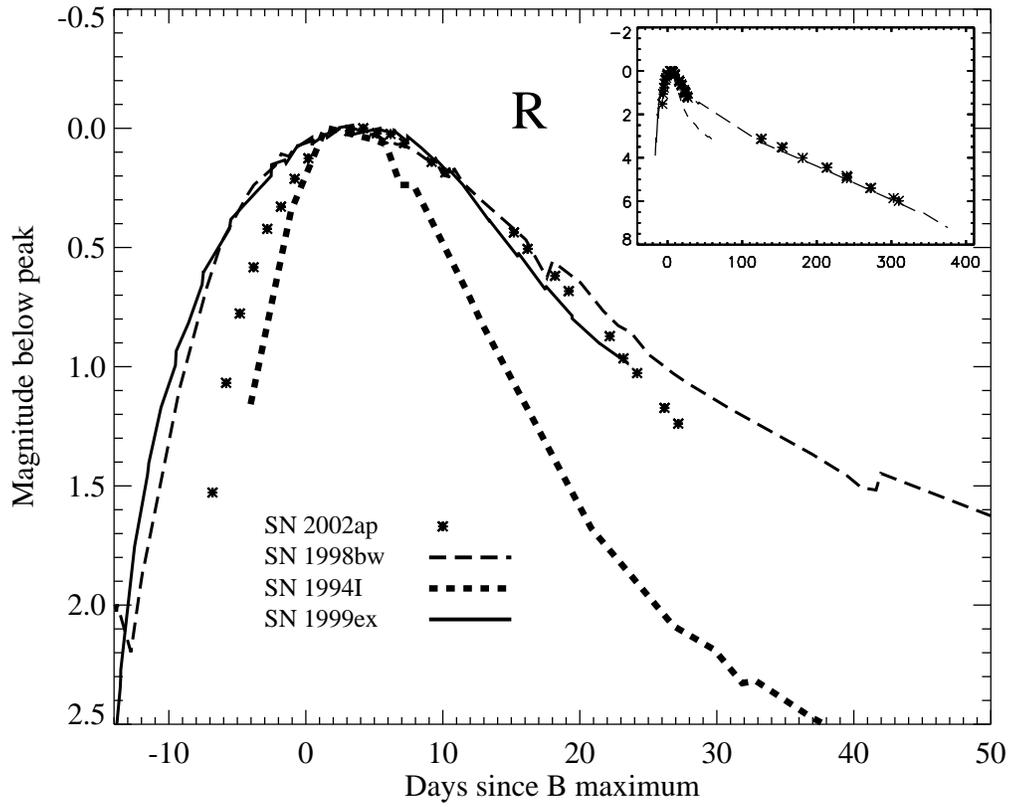}
}
}
\end{center}
\caption{Same as Figure~\ref{f:bband}, but for the $R$-band light
curve and excluding SN~1985F.}\label{f:rband}
\end{figure}

$R$-band light curves of SNe~2002ap, 1999ex, 1998bw, and 1994I are
presented in Figure~\ref{f:rband}.  For the first 6 days, $t = -7$ to
$-1$ d, SNe~2002ap and 1994I rise very quickly compared with
SNe~1998bw and 1999ex, although SN~2002ap rises at a rate somewhat
intermediate to SNe~1994I and 1998bw.  SN~1994I reaches $R$ maximum at
$t \approx 1.5$ d past $B$ maximum, while SNe~2002ap, 1999ex, and
1998bw each reach $R$ maximum at $t \approx 4$ d.

After maximum, SNe~1998bw, 1999ex, and 2002ap decline together for 16
d ($t = 4 - 20$ d).  Thereafter, SN~1998bw declines more slowly, while
SNe~2002ap and 1999ex continue their earlier decline.  Compared with
the other SNe, SN~1994I fades precipitously after $R$ maximum.

In the late phase, SN~2002ap ($0.016 \pm 0.001$ mag day$^{-1}$, $120 <
t < 310$ d) fades at a mean rate slightly faster than SN~1998bw
($0.015 \pm 0.001$ mag day$^{-1}$, $120 < t < 370$ d).

\subsubsection{$I$~Band}


\begin{figure}
\ssp
\begin{center}
\rotatebox{90}{
\scalebox{0.7}{
\plotone{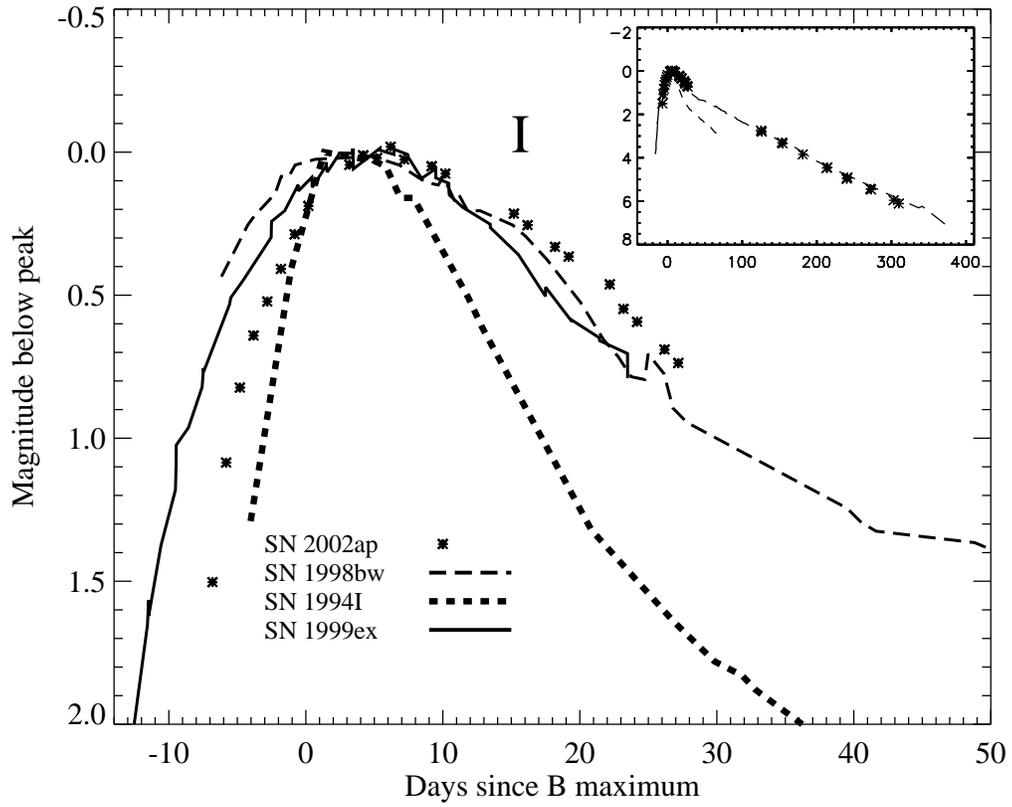}
}
}
\end{center}
\caption{Same as Figure~\ref{f:bband}, but for the $I$-band light
curve and excluding SN~1985F.}\label{f:iband}
\end{figure}

The $I$-band light curves of SNe~2002ap, 1998bw, 1999ex, and 1994I are
presented in Figure~\ref{f:iband}. As with the other bands, SNe~2002ap
and 1994I rise quickly relative to SNe~1998bw and 1999ex before
maximum, with SN~2002ap intermediate between the rest.  SN~2002ap then
reaches maximum later than all the other SNe, but is most nearly
coincident with the maximum of SN~1999ex.

In the post-maximum decline, SN~2002ap behaves like SN~1998bw until $t
\approx 17$ d, after which SN~1998bw begins declining slightly faster.
This is the opposite of what is observed in $R$, where after this
time, SN~2002ap begins to decline slightly faster than SN~1998bw.  The
light curve of SN~1999ex also exhibits a behavior similar to
SNe~2002ap and 1998bw in its post-maximum decline, but at a slightly
lower magnitude relative to the peak.  After this epoch, the light
curves of SNe~2002ap, 1998bw, and 1999ex begin to converge toward
their late-time decline phases.  As in the other bands, SN~1994I
declines very quickly after maximum and is not similar to any of the
other SNe during this time.

At late times, SNe~2002ap and 1998bw decline at mean rates of $0.019
\pm 0.001$ and $0.017 \pm 0.001$ mag day$^{-1}$ during the intervals
$120-310$ d and $120-370$ d, respectively.  Again, as in the other
bands, SN~2002ap fades slightly more quickly than SN~1998bw in the
nebular phase.

\subsection{Optical Color Curves}\label{s:color}

\begin{figure}
\ssp
\begin{center}
\rotatebox{90}{
\scalebox{0.7}{
\plotone{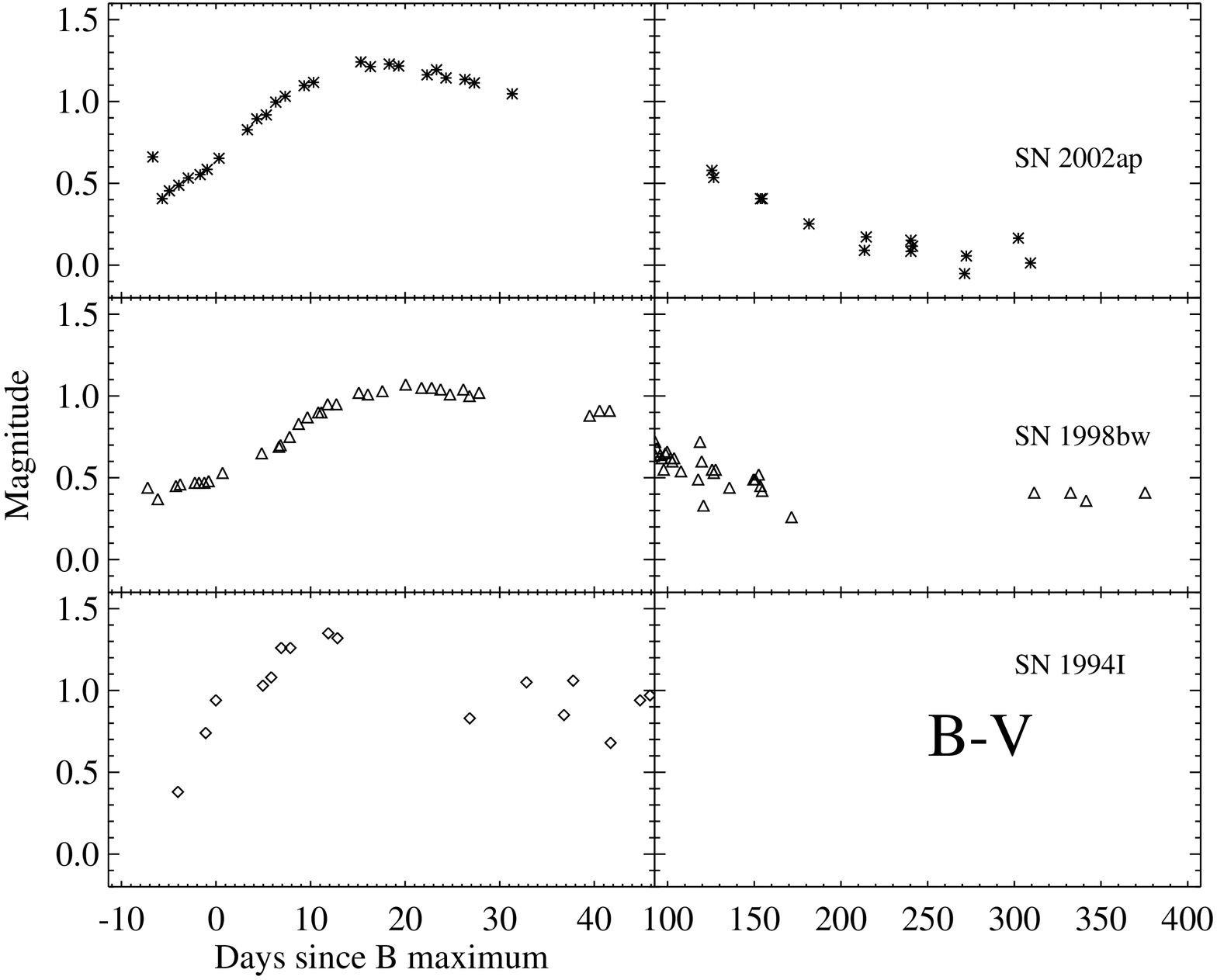}
}
}
\end{center}
\caption{The ($B - V$) color curves of SNe~2002ap, 1998bw, and
1994I. The curves are not corrected for reddening. Note the
discontinuity and the change in temporal scale at $t \approx
46$~d}\label{f:colorbv}
\end{figure}

\begin{figure}
\ssp
\begin{center}
\rotatebox{90}{
\scalebox{0.7}{
\plotone{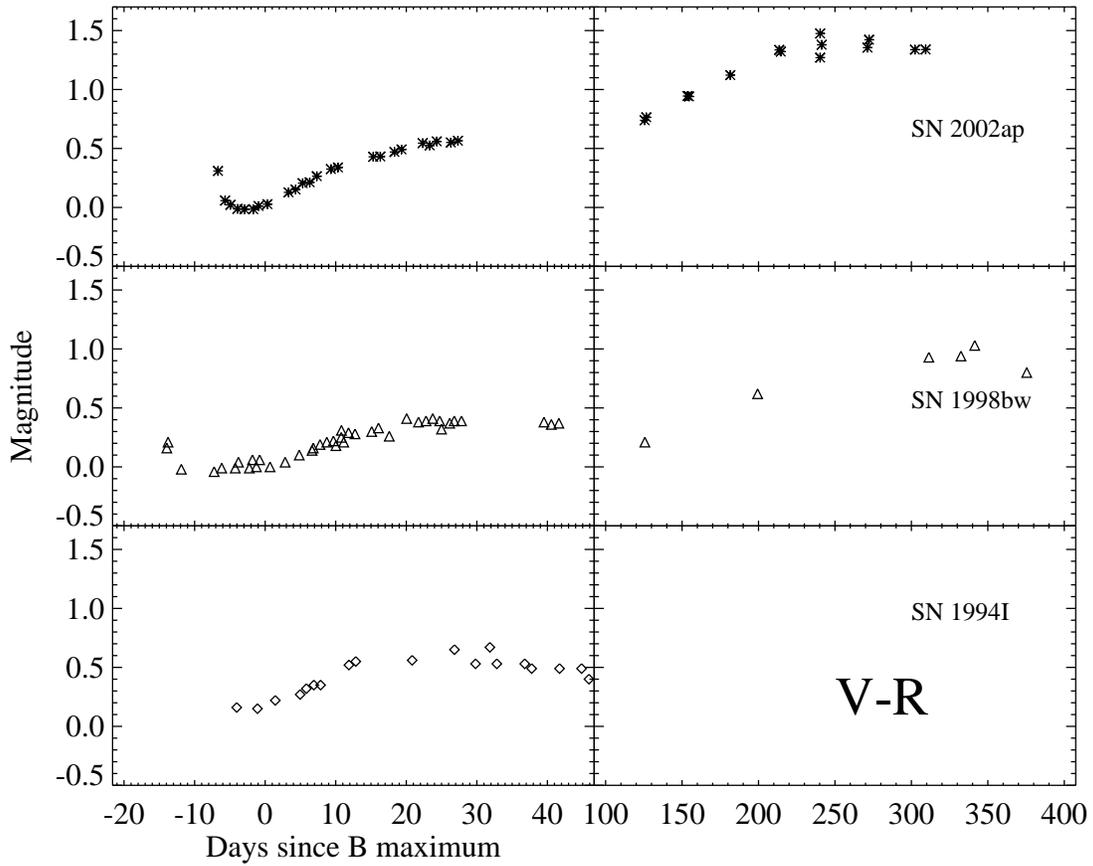}
}
}
\end{center}
\caption{The ($V - R$) color curves of SNe~2002ap, 1998bw, and
1994I. The curves are not corrected for reddening. Note the
discontinuity and the change in temporal scale at $t \approx
46$~d.}\label{f:colorvr}
\end{figure}

\begin{figure}
\ssp
\begin{center}
\rotatebox{90}{
\scalebox{0.7}{
\plotone{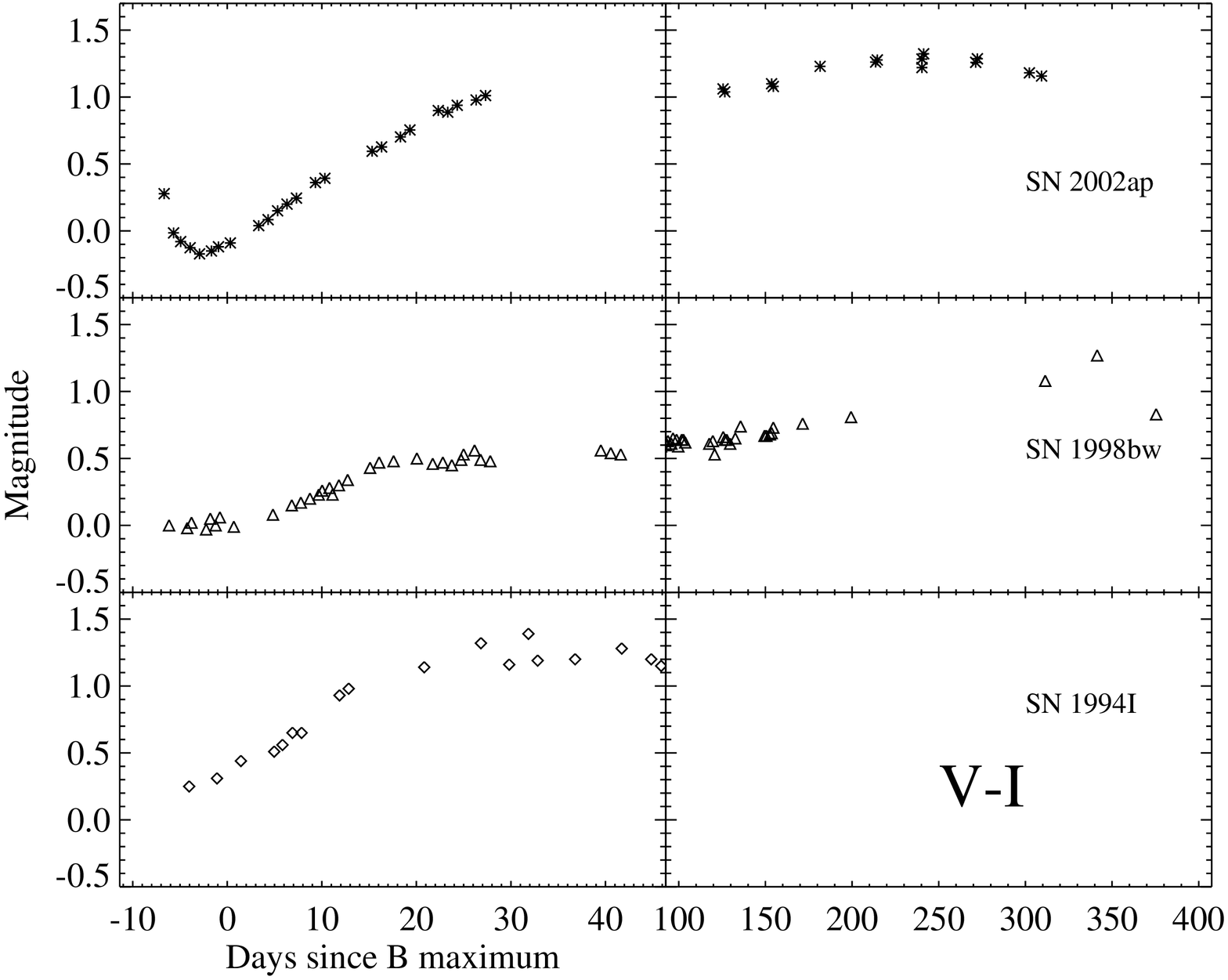}
}
}
\end{center}
\caption{The ($V - I$) color curves of SNe~2002ap, 1998bw, and
1994I. The curves are not corrected for reddening. Note the
discontinuity and the change in temporal scale at $t \approx
46$~d.}\label{f:colorvi}
\end{figure}

In Figures~\ref{f:colorbv}, \ref{f:colorvr}, and \ref{f:colorvi} we
respectively present ($B - V$), ($V - R$), and ($V - I$) color curves
of SN~2002ap and two comparisons, SNe~1998bw and 1994I.  Since the
reddenings for the comparison SNe are not well known, we present all
the curves uncorrected for reddening.

The $(B - V)$ color curves in Figure~\ref{f:colorbv} show that the
color evolution of SN~2002ap is similar to that of SN~1998bw, but
significantly different from that of SN~1994I in the early
post-maximum epoch.  However, a lack of late-time data for SN~1994I
prevents a late-time comparison with SN~2002ap.

 
All three SNe show an approximately linear increase in color at $t = 0
- 15$ d, although their slopes differ somewhat.  At later times,
SN~2002ap and SN~1998bw both exhibit a linear decrease in color after
$t = 15$ d, until $t = 30$ for SN~2002ap and until $t \approx 180$ d
for SN~1998bw.  The late-time ($B - V$) data for SN~2002ap are sparse,
but after $t \approx 200$ d SN~2002ap appears to reach a plateau phase
of nearly constant color.

The ($V - R$) color evolution of SN~2002ap differs dramatically from
the ($B - V$) evolution.  SN~2002ap and SN~1998bw appear to show
similarities in early decline rate from $t \approx -10$ d to $t = 0$ d
and in a linear increase in color between $t = 0$ d and $t \approx 30$
d.  SN~1994I also displays somewhat similar behavior during these
days.  After $t \approx 30$ d, SN~2002ap exhibits a dramatic change,
by becoming increasingly redder to a maximum at $t \approx 250$ d,
after which it appears to enter another plateau phase.  This behavior
is mimicked by SN~1998bw at these late times.  This epoch is similar
to the plateau phase in ($B - V$).


Figure~\ref{f:colorvi} shows the ($V - I$) color evolution of the SNe.
At early times, SN~2002ap shows more similarities to SN~1994I,
increasing in a similar, linear way.  Both SNe rise quite dramatically
from $t \approx 0$ d to $t \approx 30$ d, while SN~1998bw has entered
a plateau phase at this time.  Both SNe~2002ap and 1994I have a larger
change in ($V - I$) color than SN~1998bw.  At late times ($t \approx
100 - 300$ d), SN~2002ap becomes redder at a minimal rate, slightly
reversing this trend only in the last few days.  SN~1998bw displays
similar behavior, only decreasing $\sim 50$ d later.


\subsection{Overall Photometric Results}

In \ubvri\!\!\!, SN~2002ap rises quickly like the normal SN~Ic~1994I,
while it declines slowly like the high-velocity SN~Ic~1998bw and the
SN~Ib/c~1999ex, making its \ubvri light curves distinctly broader than
those of a ``normal,'' SN~1994-like SN~Ic. At late times, SN~2002ap is
again more photometrically similar to SN~1998bw than to a normal
SN~Ic.

Our photometric parameters listed in Table~\ref{t:phottable} agree
reasonably well with those found by \citet{Pandey03} and
\citet{Yoshii03}.  \citeauthor{Pandey03} report that there is a
``flattening'' of the optical light curves, where after $t \approx 30$
d the flux decline rates are \about 0.02 mag day$^{-1}$ in all
passbands until solar conjunction.  However, we did not observe this
flattening because our observations ceased approximately when
\citeauthor{Pandey03} began to see this effect.  Also,
\citet{Yoshii03} state that the rise rate before peak is almost
independent of bandpass, while the post-maximum decline rate is
steepest in $U$ and is progressively shallower through $I$.  Our data
show wavelength-dependent photometric evolution not only after
maximum, but also before (Figure~\ref{f:allcurves}).

The color evolution of SN~2002ap is quite similar to the evolution of
SN~1998bw and dissimilar from the evolution of SN~1994I.  However, the
($V - I$) colors of SNe~2002ap and 1994I are relatively similar.

\section{SPECTROSCOPY}\label{s:spec}

Our spectroscopic observations of SN~2002ap began on 2002 February
11.2, 5.4 d after $B$ maximum and 13.3 d after the time of explosion
found by \citet{Mazzali02}.  They were halted 2002 March 11 when the
supernova was in solar conjunction, but resumed on 2002 June 8.5 on a
semi-regular basis.

Data were obtained with the Kast double spectrograph \citep{Miller93}
mounted on the Lick Observatory 3-m Shane telescope, the Low
Resolution Imaging Spectrometer (LRIS; \citealt{Oke95}) mounted on the
10-m Keck~I telescope, and the Echellette Spectrograph and Imager
(ESI; \citealt{Sheinis02}) mounted on the 10-m Keck~II telescope.  The
position angle of the slit was generally aligned along the parallactic
angle to reduce differential light losses \citep{Filippenko82}.
Table~\ref{t:spec-sch} lists the journal of observations.

\begin{deluxetable}{llccccccc}
\tabletypesize{\scriptsize}
\tablewidth{0pt}
\tablecaption{Journal of Spectroscopic Observations of SN~2002ap\label{t:spec-sch}}
\tablehead{\colhead{} &
\colhead{} &
\colhead{} &
\colhead{Resolution\tablenotemark{c}} &
\colhead{Range} &
\colhead{Air} & 
\colhead{Seeing\tablenotemark{e}} &
\colhead{Exposure} &
\colhead{} \\ 
\colhead{Day\tablenotemark{a}} &
\colhead{UT Date} &
\colhead{Telescope\tablenotemark{b}} &
\colhead{(\AA)} &
\colhead{(\AA)} &
\colhead{Mass\tablenotemark{d}} &
\colhead{(arcsec)} &
\colhead{(s)} &
\colhead{Observer\tablenotemark{f}}}

\startdata
5.4   & 2002 Feb 11.2 & Kast  & 12 & 3300-10400 & 1.7 & 3.0 & 400  & AF, RC \\
8.4   & 2002 Feb 14.2 & LRISp & 12 & 3800-10100 & 1.6 & 2.0 & 900  & AF, AB \\
8.5   & 2002 Feb 14.3 & LRISp & 12 & 3930-8830  & 2.2 & 2.5 & 2400 & AF, AB \\
8.5   & 2002 Feb 14.3 & LRISp & 12 & 3800-10100 & 3.3 & 4.5 & 1100 & AF, AB \\
15.4  & 2002 Feb 21.2 & Kast  & 6  & 3300-10400 & 3.9 & 3.5 & 1200 & WL, RC \\
29.4  & 2002 Mar 7.2  & LRISp & 12 & 3900-8800  & 2.7 & 1.5 & 1200 & AF, DL, EM \\
29.5  & 2002 Mar 7.3  & LRISp & 12 & 3900-8800  & 3.7 & 2.0 & 940  & AF, DL, EM \\
33.3  & 2002 Mar 11.1 & Kast  & 7  & 3300-10400 & 3.0 & 2.5 & 1200 & WL, RC \\
122.7 & 2002 Jun 8.5  & Kast  & 6  & 3290-10275 & 3.5 & 7.0 & 1200 & AF, RC, RF \\
131.7 & 2002 Jun 17.5 & Kast  & 8  & 4280-7040  & 2.3 & 4.0 & 500  & AF, RC, RF \\
155.7 &	2002 Jul 11.5 & Kast  & 6  & 3150-10400 & 1.4 & 1.5 & 1400 & AF, RC, RF \\
184.6 &	2002 Aug 9.4  & Kast  & 7  & 3160-10400 & 1.2 & 2.5 & 1200 & AF, RF, MP, BS \\
237.3 & 2002 Oct 1.1  & Kast  & 7  & 3240-10400 & 1.4 & 2.0 & 1800 & AF, RF \\
241.7 & 2002 Oct 8.5  & LRIS  & 6  & 3200-9300  & 1.2 & 1.0 & 600  & AF, RC \\
273.6 & 2002 Nov 6.4  & ESI   & 1.5\tablenotemark{g} & 4000-10100 & 1.0 & 1.5 & 600 & AF, SJ, RC \\
335.6 & 2003 Jan 7.4  & LRIS  & 6  & 3215-9280  & 2.6 & 2.0 & 600  & AF, RC \\
386.4 & 2003 Feb 27.2 & LRIS  & 6  & 3200-9390  & 1.9 & 1.0 & 700  & AF, RC
\enddata 

\tablecomments{LRISp observations were performed in dual-beam mode
with a 1\farcs5 slit and a D680 dichroic separating the blue and red
parts of the spectrum; an OG570 order-blocking filter was also
inserted on the red side.  LRIS observations used a D560 dichroic and
a 1\farcs0 slit.  ESI observations used a 1\farcs0 slit.  A D550
dichroic and either a 3\farcs0 or 2\farcs0 slit were used for all Lick
observations.}

\tablenotetext{a}{Day since $B$-band maximum, 2002 Feb 5.8 (HJD
2,452,311.3).  To calculate the day since estimated date of explosion,
add 7.9 \citep{Mazzali02}.}

\tablenotetext{b}{Kast = Lick 3-m/Kast double spectrograph
\citep{Miller93}; LRIS(p) = Keck-I 10-m/Low Resolution Imaging
Spectrometer (\citealt{Oke95}; LRIS, ``p'' denotes polarimeter
mounted); ESI = Keck-II 10-m/Echellette Spectrograph and Imager
(\citealt{Sheinis02}; ESI).}

\tablenotetext{c}{Approximate spectral resolution derived from
night-sky lines.}

\tablenotetext{d}{Beginning airmass for each set of observations.}

\tablenotetext{e}{Average value of the full width at half maximum
(FWHM) of the spatial profile for each set of observations, rounded to
the nearest 0\farcs5.}

\tablenotetext{f}{AB = Aaron Barth, AF = Alex Filippenko, BS = Brandon
Swift, DL = Doug Leonard, EM = Ed Moran, MP = Marina Papenkova, RC =
Ryan Chornock, RF = Ryan Foley, SJ = Saurabh Jha, WL = Weidong Li.}

\tablenotetext{g}{Resolution at 6000 \AA.  ESI resolution is \about
75\kms\ across the entire wavelength range.}

\end{deluxetable}

All data were reduced using standard techniques as described by
\citet{Li01} and references therein.  Flatfields for the red CCD were
taken at the position of the object to reduce near-IR fringing
effects.  The spectra were corrected for atmospheric extinction and
telluric bands \citep{Bessell99,Matheson00c}, and then flux calibrated
using standard stars observed at similar airmass on the same night as
the SN.  All spectra in this paper have been deredshifted by the
NED\footnote{The NASA/IPAC Extragalactic Database (NED) is operated by
the Jet Propulsion Laboratory, California Institute of Technology,
under contract with the National Aeronautics and Space
Administration.} redshift of each SN's host galaxy ($z = 0.002192$ for
SN~2002ap).  The SN~2002ap spectra were dereddened by the overall
reddening of $E(B-V) = 0.079$ mag as determined in
Section~\ref{ss:red}.  The zero-velocity \ion{Na}{1} D lines in our
ESI spectrum have an equivalent width of \about 0.4~\AA.  By the
relationship proposed in \citet{Barbon90}, this corresponds to a
Galactic reddening of $E(B-V) \approx 0.1$ mag, which is consistent
with the value we have determined.  All other spectra were also
dereddened by the appropriate Galactic value from \citet{Schlegel98}.

\begin{figure}
\ssp
\begin{center}
\rotatebox{90}{
\scalebox{1}{
\plotone{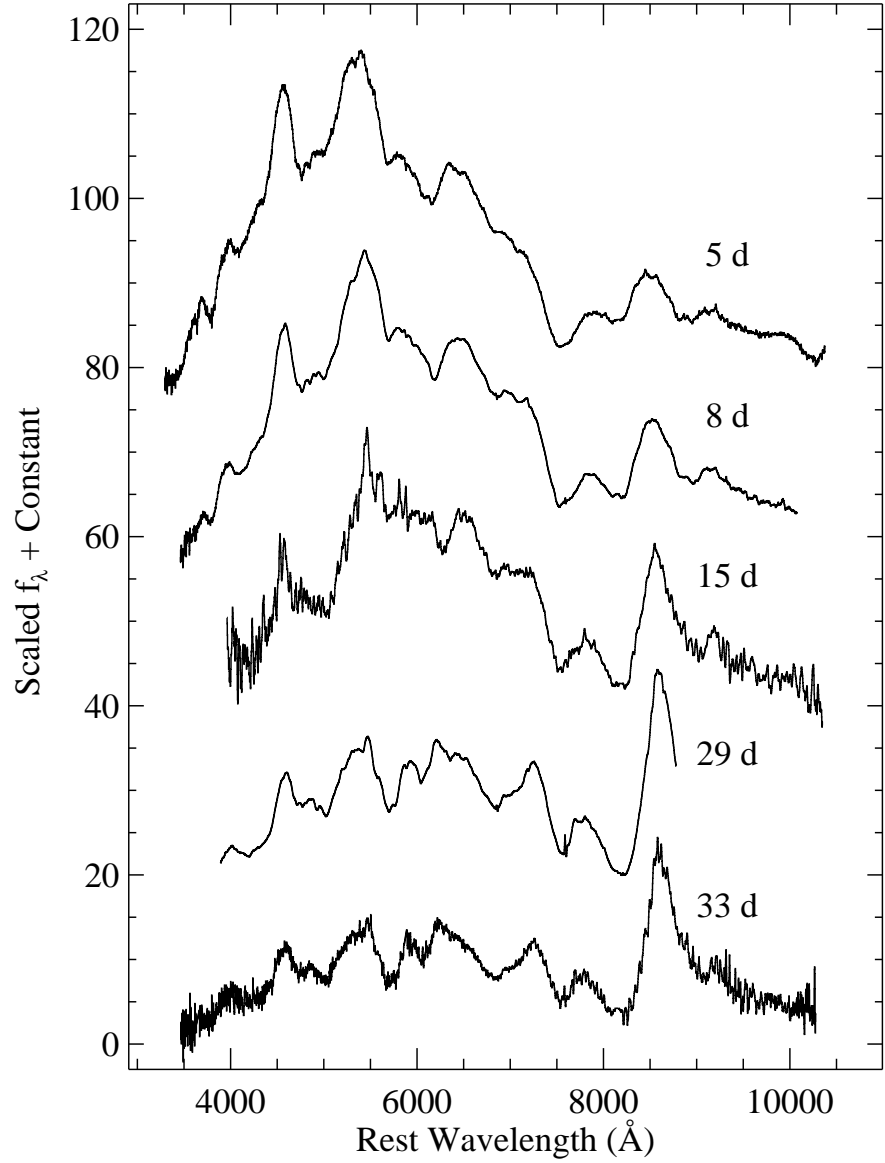}
}
}
\end{center}
\caption{Spectra of SN~2002ap for $t < 40$ d relative to $B$~maximum.
For a clear comparison, all spectra were scaled such that the
\ion{O}{1} $\lambda 7774$ absorption feature is roughly equal to that
of the $t = 5$ d spectrum and then shifted vertically by arbitrary
amounts.}
\label{f:early_spectra}
\end{figure}

\begin{figure}
\ssp
\begin{center}
\rotatebox{90}{
\scalebox{0.7}{
\plotone{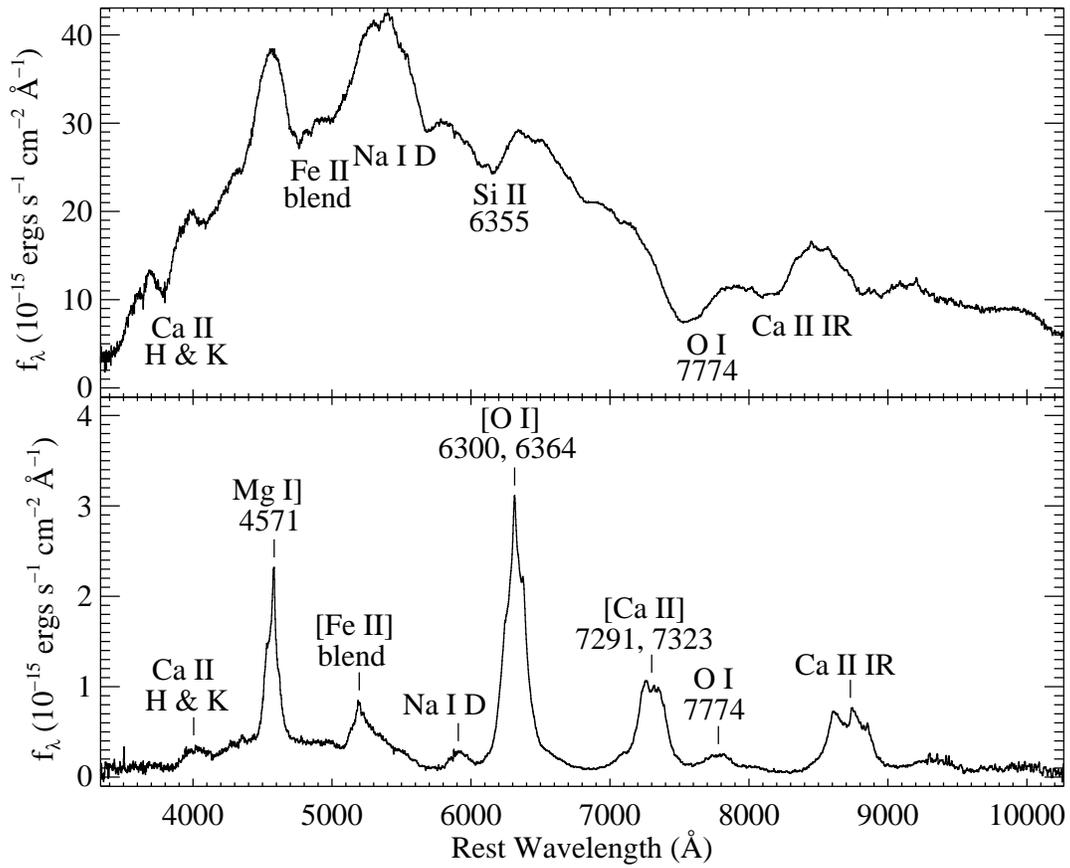}
}
}
\end{center}
\caption{Spectra of SN~2002ap at 5 and 156 d past $B$~maximum.  The
identifications are either the result of modeling (see
Figure~\ref{f:latefit}) or are common SN features.}
\label{f:identifications}
\end{figure}

We display our early-time SN~2002ap spectra in
Figure~\ref{f:early_spectra}.  Line identifications for the early-time
and late-time spectra are shown in Figure~\ref{f:identifications}.
With a lack of hydrogen, helium, and strong $6150$~\AA\ \ion{Si}{2}
features, SN~2002ap is classified as a SN~Ic
\citep{Kinugasa02,Meikle02,Gal-Yam02b}.  The high velocities present
at early times indicate that the object produced an especially large
energy per unit mass of the ejecta.  The broad features at \about 4500
and \about 5400 \AA\ resemble those in the spectra of SN~1998bw, the
prototypical ``hypernova'' associated with GRB 980425
\citep{Galama98}.  The type Ic classification, large velocities, and
spectral similarity to SN~1998bw caused many to classify SN~2002ap as
a ``hypernova.''


\citet{Mazzali02} attributed the feature at \about 3800~\AA\ in the
early-time spectra of SN~2002ap (see Figure~\ref{f:identifications})
to \ion{Ca}{2} H \& K.  The broad feature from \about 4700 to \about
5400~\AA\ is caused by an iron blend.  The small feature at 5700~\AA\
was attributed to either \ion{Na}{1} or \ion{He}{1}.  However, for
reasons discussed below, the progenitor of SN~2002ap probably had very
little helium; hence, the feature is most likely \ion{Na}{1}.  The
features from \about 6000 to \about 6400~\AA, from \about 7400 to
\about 8000~\AA, and from \about 8100 to \about 8700~\AA\ are
\ion{Si}{2}, \ion{O}{1}, and \ion{Ca}{2}, respectively.


\begin{figure}
\ssp
\begin{center}
\rotatebox{90}{
\scalebox{1}{
\plotone{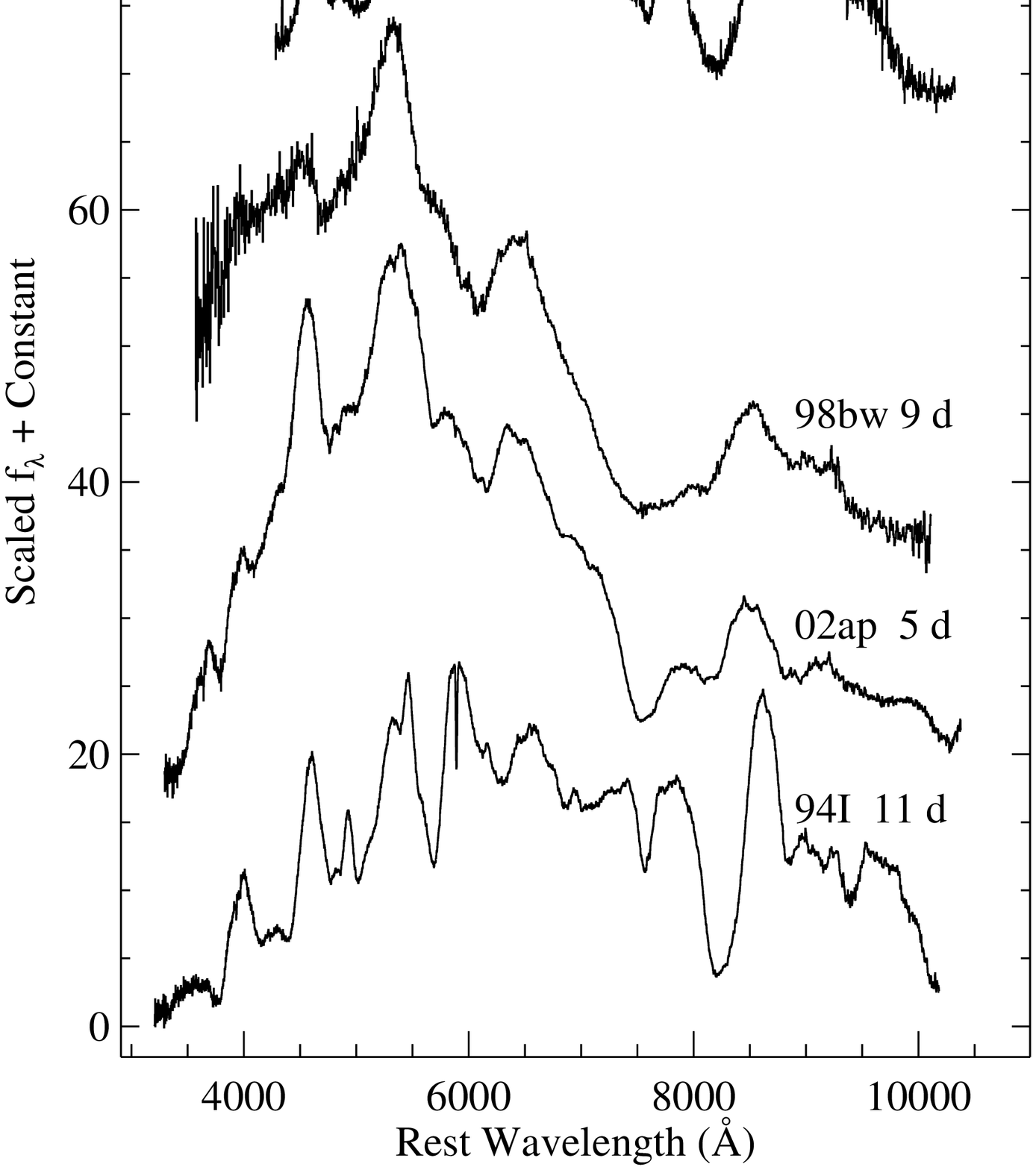}
}
}
\end{center}
\caption{A comparison of SNe~2002ap, 1998bw \citep{Stathakis00},
1997ef \citep{Matheson01}, and 1994I \citep{Filippenko95} at early
times.  The phases marked are relative to $B$~maximum.  The spectra
are scaled such that the \ion{O}{1} $\lambda7774$ absorption feature
is roughly equal to that of the SN~2002ap spectrum and then shifted
vertically by arbitrary amounts.  The narrow emission lines in the
spectra of SN~1998bw and SN~1997ef are from superimposed \ion{H}{2}
regions.}
\label{f:early_compare}
\end{figure}

In Figure~\ref{f:early_compare}, we present an early-time spectral
comparison of SN~2002ap with the ``hypernovae'' SNe~1998bw and 1997ef
as well as the normal SN~Ic~1994I at early times.  SN~2002ap is very
similar to SN~1998bw, somewhat similar to SN~1997ef (over the very
limited available wavelength range), and less similar to SN~1994I.
Specifically, SN~1994I has narrower lines and much better-developed
P-Cygni profiles for the \ion{Na}{1} D and \ion{Ca}{2} near-IR triplet
lines, with very strong absorption components.  The size of the
\ion{Ca}{2} triplet relative to other features can be used to identify
the age of a SN~Ic \citep{Matheson00c}.  This suggests that SN~2002ap
evolved more slowly than SN~1994I.

Figure~\ref{f:early_compare} also shows that certain spectral features
are washed out with higher velocity.  Specifically, SN~1994I has a
significant dip in the spectra around 5400 \AA.  This feature is also
present in the SN~2002ap spectrum, although not as prominent, but is
not visible in the SN~1998bw spectrum.  This washing out is indicative
of higher energy per unit mass.


\begin{figure}
\ssp
\begin{center}
\rotatebox{90}{
\scalebox{0.7}{
\plotone{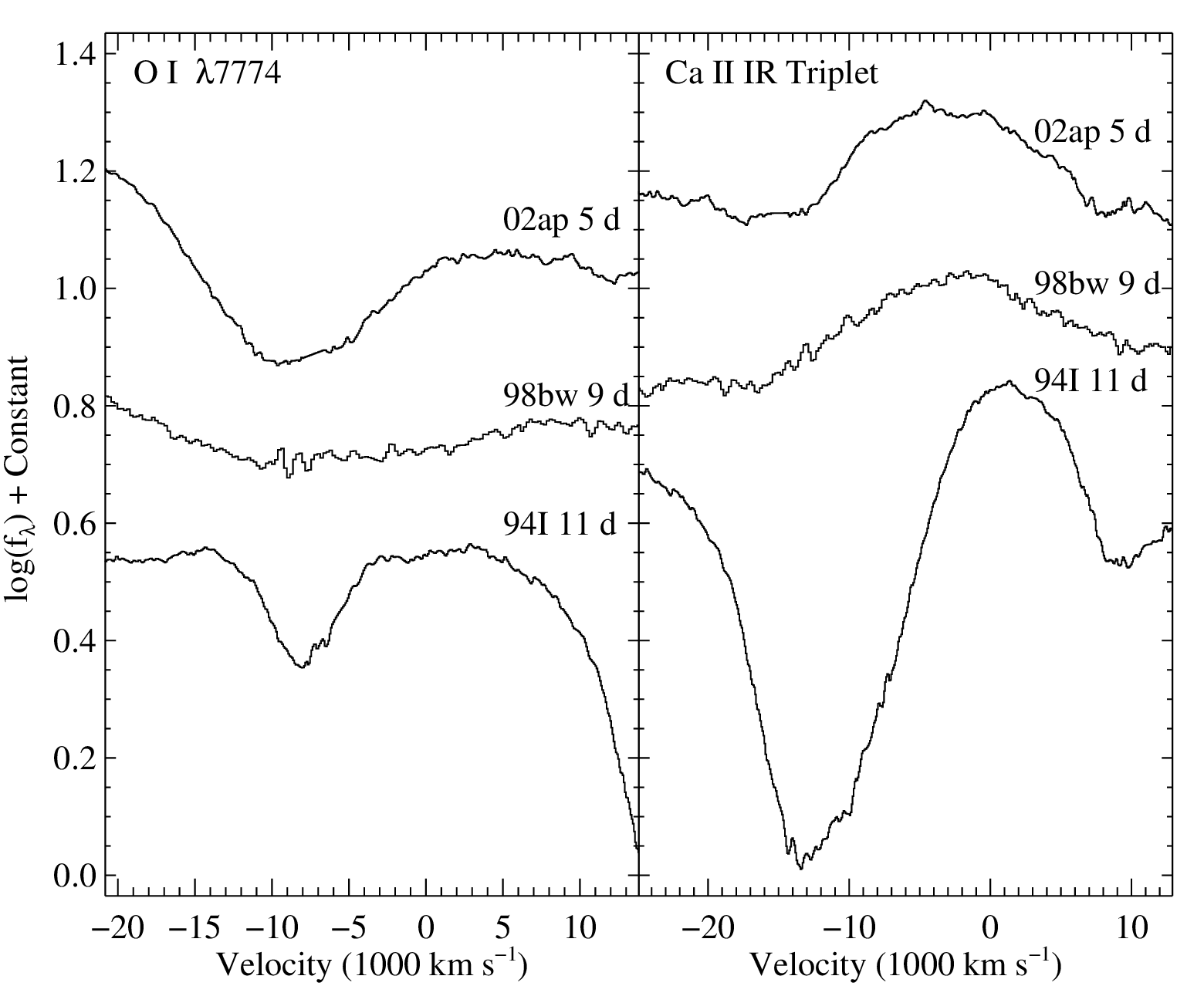}
}
}
\end{center}
\caption{Spectra of SNe~2002ap, 1998bw, and 1994I on a velocity
scale. The zero velocity corresponds to the \ion{O}{1} $\lambda7774$
line and the $gf$-weighted \ion{Ca}{2} $\lambda 8579$ \citep{Thomas03}
in the rest frame.}
\label{f:velocity}
\end{figure}

Comparisons of velocity widths of the \ion{O}{1} $\lambda7774$ line
and the \ion{Ca}{2} IR triplet for SNe~2002ap, 1998bw, and 1994I are
shown in Figure~\ref{f:velocity}.  For SN~2002ap, \citet{Filippenko02}
measured a photospheric velocity from \ion{O}{1} $\lambda7774$ of
\about 9000 \kms\ at 5 d after $B$ maximum.  This is consistent with
measurements from \citet{Gal-Yam02a} and similar to SN~1998bw at the
same epoch.  The figure shows that SN~1998bw has a slightly lower
photospheric velocity at 9 d after $B$ maximum than SN~2002ap at 5 d
past $B$ maximum.  Photospheric velocity decreases rapidly at early
times, but slows about a week after maximum.  The large photospheric
velocity seen in SN~2002ap suggests that SN~2002ap is more energetic
per unit mass of ejecta than normal SNe~Ic such as SN~1994I.

>From both the photometric light curves and the spectral comparisons,
it is evident that SN~2002ap ages more quickly than SN~1998bw or
SN~1997ef.  The faster aging and the lower velocities derived from the
line widths indicate that SN~2002ap is less energetic than SN~1998bw
\citep{Kinugasa02}.  Likewise, the slower aging relative to SN~1994I
suggests that SN~2002ap is more energetic than normal SN~Ic (for
energetic comparisons, see Table~\ref{t:params}).


\begin{figure}
\ssp
\begin{center}
\rotatebox{90}{
\scalebox{1}{
\plotone{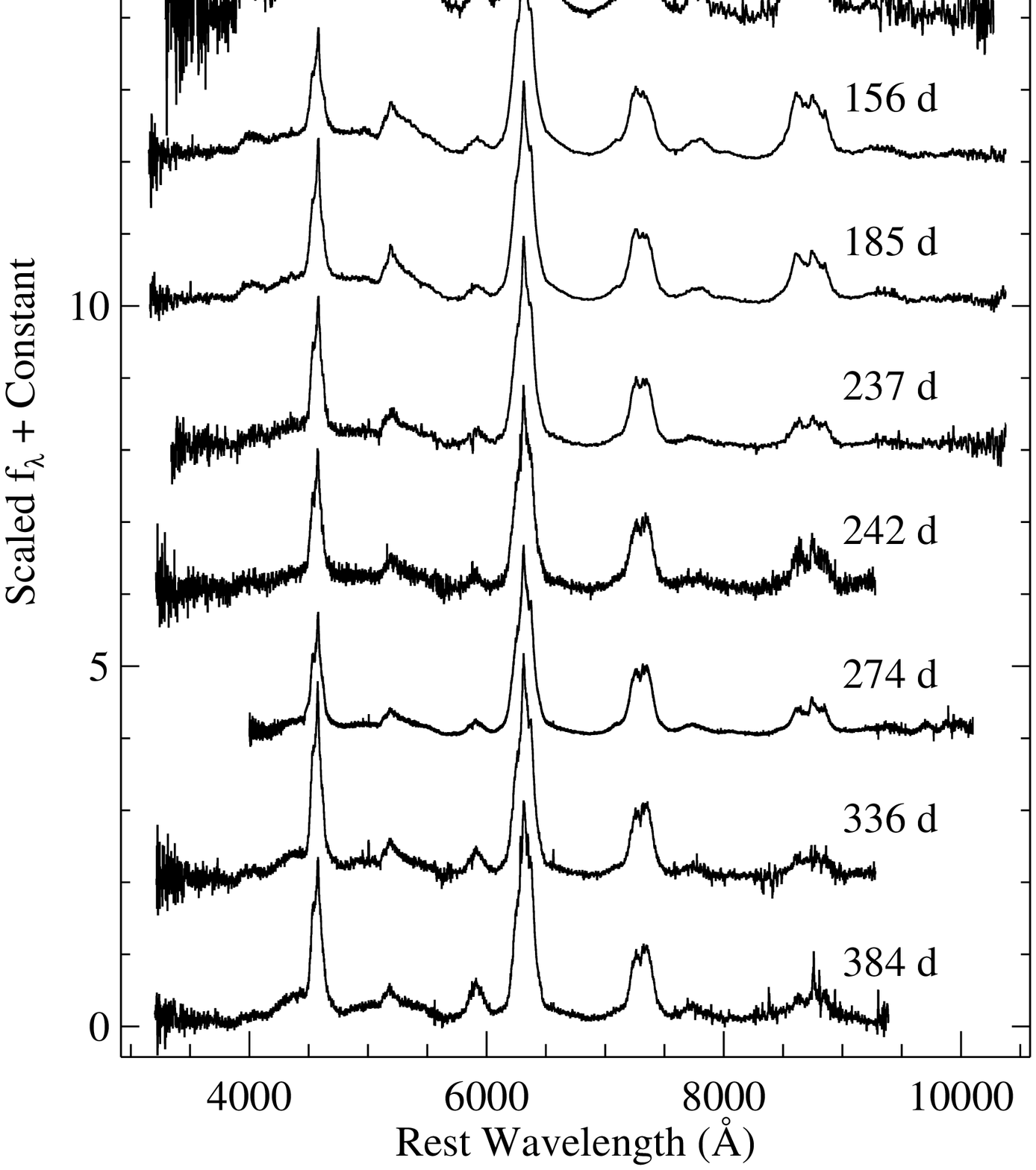}
}
}
\end{center}
\caption{Late-time spectra of SN~2002ap.  The phases are marked
relative to $B$~maximum.  Our $t = 123$ and $t = 132$ d spectra were
averaged to reduce noise.  The spectra are scaled such that the height
of the [\ion{Ca}{2}] $\lambda \lambda 7291$, 7329 line in each
spectrum is equal to that of the $t = 123 + 132$ d spectrum and then
shifted vertically by arbitrary amounts.  The blue-side shutter of
LRIS was not functional when the 336 d spectrum was obtained, and thus
the scaling between features redward of 5500~\AA\ and blueward of
5500~\AA\ may be somewhat erroneous.}\label{f:late_spectra}
\end{figure}

The nebular spectra of SN~2002ap shown in
Figures~\ref{f:identifications} and \ref{f:late_spectra} have
incredibly strong [\ion{O}{1}] $\lambda \lambda 6300$, 6364 and
\ion{Mg}{1}] $\lambda 4571$ emission.  The \ion{Mg}{1}] line grows
over time with respect to the [\ion{Ca}{2}] $\lambda \lambda 7291$,
7324 doublet and the [\ion{O}{1}] doublet.  In our latest spectrum
(taken on 2003 February 27, 386 d past maximum), the \ion{Mg}{1}]
integrated flux is 40\% the [\ion{O}{1}] integrated flux. Since we see
farther into the core of the progenitor with time, we are probably
peering directly into the Mg-O layer.  It is expected that
observations at later times will show an even larger \ion{Mg}{1}] to
[\ion{O}{1}] ratio.

\begin{figure}
\ssp
\begin{center}
\rotatebox{90}{
\scalebox{1}{
\plotone{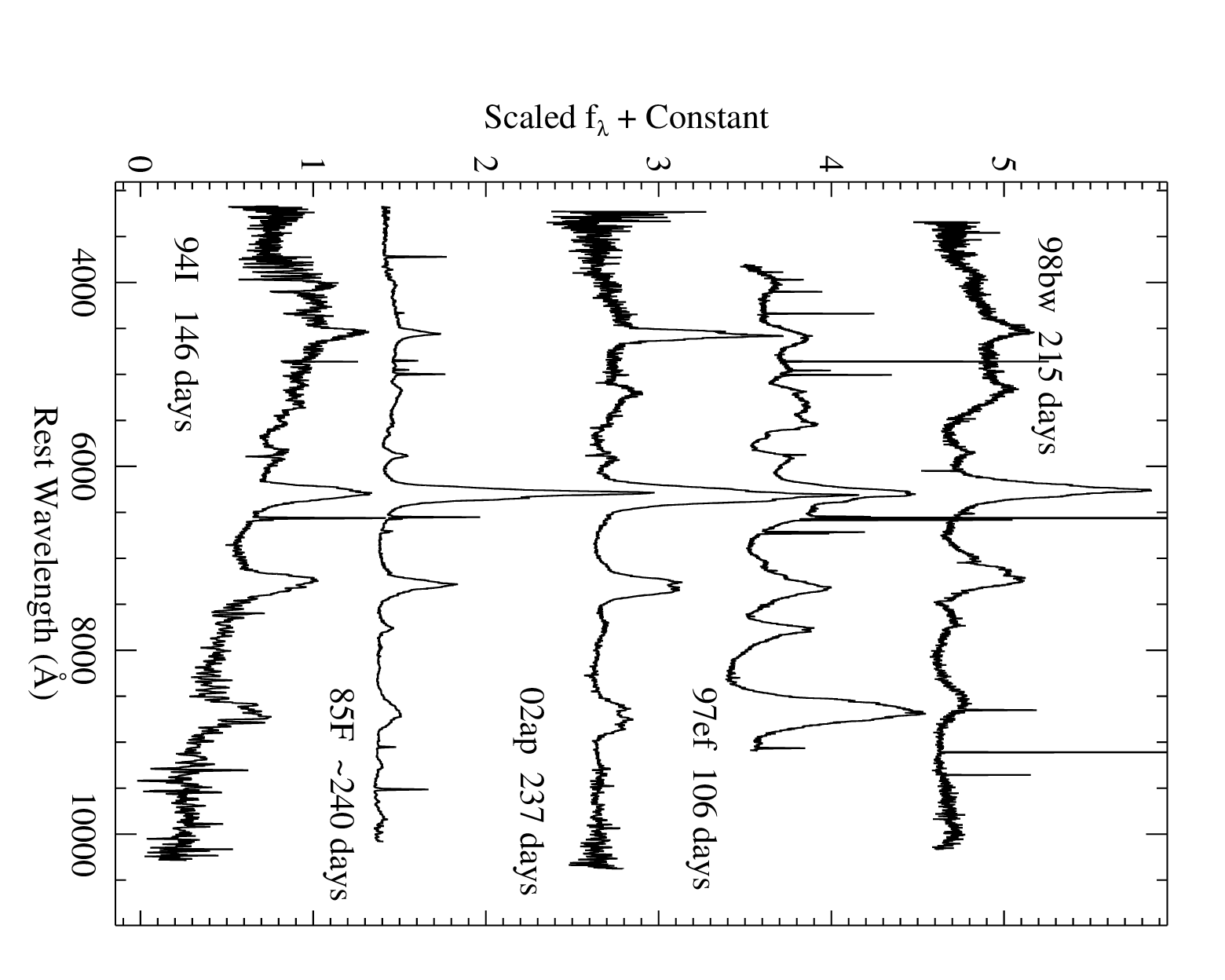}
}
}
\end{center}
\caption{A comparison of spectra of SNe~2002ap, 1998bw
\citep{Sollerman00}, 1997ef \citep{Matheson01}, 1994I
\citep{Filippenko95}, and 1985F \citep{Filippenko86} during the
nebular phase.  The times marked are relative to $B$~maximum.  The
spectra are scaled such that the [\ion{Ca}{2}] $\lambda \lambda 7291$,
7324 emission feature of each spectrum is roughly equal in height and
then shifted vertically by arbitrary amounts.  The spectrum of
SN~1994I is heavily contaminated by superposed stars.  Narrow emission
lines are in all cases due to \ion{H}{2} regions.}
\label{f:late_compare}
\end{figure}

As seen in Figure~\ref{f:late_compare}, the features of SNe~2002ap,
1998bw, 1997ef, 1985F, and 1994I are all similar, but the relative
strengths of some emission lines differ.  Relative to [\ion{Ca}{2}]
$\lambda \lambda 7291$, 7324, the \ion{Mg}{1}] $\lambda 4571$ emission
is strongest in SN~2002ap, with the other SNe having about the same
line strength.  The other striking difference is the very strong
\ion{O}{1} $\lambda 7774$ and Ca IR triplet in the spectrum of
SN~1997ef. In contrast, the [\ion{O}{1}] $\lambda \lambda 6300$, 6364
and [\ion{Ca}{2}] $\lambda \lambda 7291$, 7324 have about the same
strength in SN~1997ef as in the other SNe.  The stronger permitted
lines relative to forbidden lines is likely due to denser radiating
material in SN~1997ef.  Indeed, the excess emission near 5400~\AA\ may
be \ion{Fe}{2} emitted by dense clouds, as in SN~1987F
\citep{Filippenko89}.

Another prominent feature is the near-IR \ion{Ca}{2} triplet.  This
blend decreases in strength over time relative to [\ion{Ca}{2}].
\citet{Matheson00b, Matheson00c} found that the strength of the
\ion{Ca}{2} triplet depends heavily on age for SNe~Ib/c, and should
decrease in strength over time, consistent with our spectra.

The feature covering 5200 to 5600~\AA\ is probably a mixture of Mg and
Fe lines.  From modeling, we have seen that the shoulder of the
feature may be [\ion{Fe}{2}], whereas the emission at 5190~\AA\ is
probably \ion{Mg}{1}].  In some SNe with very dense ejecta, however,
\citet[][SN~1987F]{Filippenko89} suggests that \ion{Fe}{2} lines
dominate.

\begin{figure}
\ssp
\begin{center}
\rotatebox{90}{
\scalebox{1}{
\plotone{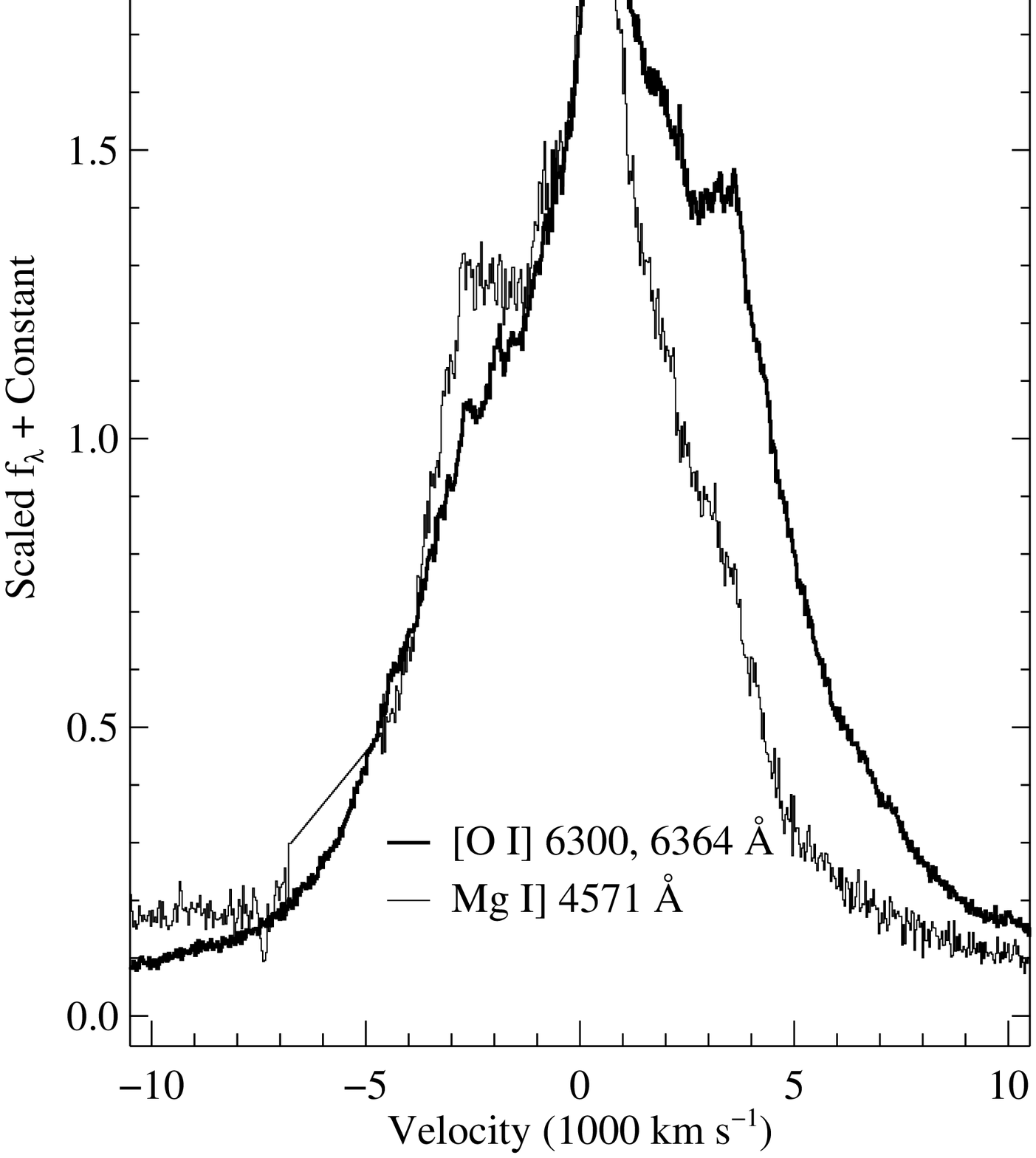}
}
}
\end{center}
\caption{The \ion{Mg}{1}] $\lambda 4571$ and [\ion{O}{1}] $\lambda
\lambda 6300$, 6364 line profiles of SN~2002ap at 274 d past $B$
maximum.  The narrow emission components are redshifted at \about 555
\kms\ relative to M74.}
\label{f:lineprofile}
\end{figure}

The line profiles of \ion{Mg}{1}] $\lambda 4571$ and [\ion{O}{1}]
$\lambda \lambda 6300$, 6364 seen in Figures~\ref{f:identifications}
and \ref{f:late_spectra} are very sharp.  Figure~\ref{f:lineprofile}
shows these lines in detail.  The lines have a narrow component at
\about 555 \kms\ relative to the systemic velocity of M74 on top of a
slightly broader base, as first seen by \citet{Leonard02}.  Our ESI
spectrum has a high resolution of \about 75 \kms, confirming the
results of \citet{Leonard02}.  These profiles can be caused either by
low-velocity gas clouds near the center of the forming nebula or from
material lost earlier or expelled by the progenitor that is now in the
circumstellar environment of SN~2002ap \citep[for example,
SN~1999cq;][]{Matheson00a}.  The other strong emission lines
([\ion{Ca}{2}] $\lambda \lambda 7291$, 7329; \ion{Ca}{2} IR triplet)
might weakly show this profile; however, this is uncertain due to line
blending.


Table~\ref{t:line_strengths} lists the relative line strengths for
each nebular epoch of SN~2002ap.  To compute these values, first we
fit a cubic spline to a local continuum around each line.  After
subtracting the continuum, we integrated the flux in the lines.  Since
the examined lines are rather broad, other lines may add significant
flux to the wings (especially for the \ion{Ca}{2} IR triplet;
\citealt{Filippenko86} showed that [\ion{C}{1}] $\lambda 8727$
contributes to this feature).  Furthermore, to avoid obvious adjacent,
close lines, we made somewhat arbitrary determinations of the
endpoints of each line.  Changing these boundaries by a few \AA\ will
change the integrated flux slightly.  Although this method is somewhat
inaccurate, it does show that the \ion{Mg}{1}] increases in strength
and the near-IR Ca triplet decreases in strength over time relative to
[\ion{Ca}{2}].  SN~2002ap had a Mg/O ratio comparable to that of
SN~1994I, but larger than that of SN~1998bw.


\begin{deluxetable}{ccccccc}
\tabletypesize{\footnotesize}
\tablewidth{0pt}
\tablecaption{Integrated Flux in Lines\label{t:line_strengths}}
\tablehead{
\colhead{} &
\colhead{Days Past} &
\colhead{\ion{Mg}{1}]} &
\colhead{[\ion{O}{1}]} &
\colhead{[\ion{Ca}{2}]} &
\colhead{Near-IR \ion{Ca}{2}} &
\colhead{Mg/O} \\
\colhead{Supernova} &
\colhead{$B$ Maximum} &
\colhead{(4571~\AA)} &
\colhead{(6300, 6364~\AA)} &
\colhead{(7291, 7324~\AA)} &
\colhead{(8498, 8542, 8662~\AA)} &
\colhead{Ratio}}

\startdata
SN~2002ap & $122.7 + 131.7$ & 0.66 & 2.75 & 1 & 1.85 & 0.24 \\
SN~2002ap & 155.7           & 0.66 & 2.53 & 1 & 1.43 & 0.26 \\
SN~2002ap & 184.6           & 0.74 & 2.35 & 1 & 1.01 & 0.32 \\
SN~2002ap & 237.3           & 0.80 & 2.29 & 1 & 0.63 & 0.35 \\
SN~2002ap & 241.7           & 0.72 & 2.17 & 1 & 0.91 & 0.33 \\
SN~2002ap & 273.6           & 0.63 & 2.01 & 1 & 0.66 & 0.31 \\
SN~2002ap & 335.6           & 1.00 & 2.29 & 1 & 0.45 & 0.43 \\
SN~2002ap & 386.4           & 0.87 & 2.19 & 1 & 0.46 & 0.40 \\
SN~1998bw & 215             & 0.23 & 1.47 & 1 & 0.41 & 0.15 \\
SN~1994I  & 146             & 0.26 & 0.97 & 1 & 0.84 & 0.26 \\
SN~1985F & \about 240       & 0.26 & 3.15 & 1 & 0.56 & 0.08
\enddata

\tablecomments{All values given relative to the integrated flux of the
[\ion{Ca}{2}] $\lambda \lambda 7921$, 7324 emission of each spectrum.
The blue-side shutter of LRIS was not functional when the 336~d
spectrum was obtained, and thus the scaling between features redward
of 5500~\AA\ and blueward of 5500~\AA\ may be somewhat erroneous. The
calibration of the 242~d spectrum may be faulty at the longest
wavelengths, making the relative flux of the Ca~II near-IR triplet
especially uncertain.}

\end{deluxetable}

Comparisons to other SNe~Ic at late times confirm that SN~2002ap has
the strongest oxygen and magnesium lines of any SN yet published.  The
best comparison is to SN~1998bw, which has strong [\ion{O}{1}]
emission, but its \ion{Mg}{1}] line is much weaker relative to either
[\ion{O}{1}] or \ion{Ca}{2} than in SN~2002ap.

\section{MODELING}\label{s:model}

We modeled the nebular spectra of SN~2002ap using a non-LTE nebular
code \citep{Mazzali01}. (Preliminary results in agreement with those
presented here are given by \citealt{Mazzali03}.)  The code computes
$\gamma$-ray deposition in a nebula of constant density and
homogeneous composition, and produces a synthetic spectrum by
balancing the collisional heating resulting from the $\gamma$-ray
deposition and the cooling via line emission.  Emission is mostly in
forbidden lines, but some permitted transitions (e.g., \ion{Ca}{2})
are also strong if the conditions are favorable.  The model is
characterized by the outer velocity of the nebula, which is derived
essentially by fitting the width of the emission lines; the mass of
$^{56}$Ni, which provides the heating; and the masses of the various
elements required to reproduce the observed emission lines.  The mass
of $^{56}$Fe is derived from the decay of $^{56}$Ni. Since Fe lines
are observed, all masses can be determined uniquely for a given
distance and reddening.

The nebular spectra of SN~2002ap are similar to those of SN~1998bw.
The strongest feature is [\ion{O}{1}] $\lambda \lambda 6300$, 6364.
However, the blend of [\ion{Fe}{2}] lines observed at \about 5200~\AA\
is much weaker in SN~2002ap than in SN~1998bw.  Qualitatively, this is
a confirmation of the suggestion that SN~2002ap synthesized much less
$^{56}$Ni than SN~1998bw.  Unfortunately, this makes it difficult to
determine the mass of $^{56}$Ni with high precision, as the
[\ion{Fe}{2}] feature is typically observed with a small
signal-to-noise ratio and blends with other lines may affect the
measured flux.  Also, as noted above, the \ion{Mg}{1}] line is much
stronger, relative to the [\ion{O}{1}] line, in SN~2002ap.

In order to increase the accuracy of our diagnostics, we therefore
modeled three spectra, observed at different epochs.  For all spectra
we find that the outer velocity of the nebula is \about 5500\kms, and
the $^{56}$Ni mass is \about $0.09~M_{\odot}$.  The ejecta mass
contained within the outer velocity is \about $1.5~M_{\odot}$.  This
mass is consistent with the corresponding value derived from the
explosion model used by \citet{Mazzali02}.  The model ejecta extend to
much larger velocities. However, material at the highest velocities is
too thin to contribute to net emission. Hence, the estimated mass here
is a lower limit to the total ejecta mass.

Approximately $0.6~M_{\odot}$ of the ejecta mass is O, while Si and S
contribute a combined mass of \about $0.5~M_{\odot}$, and the mass of
C is \about $0.2~M_{\odot}$.  The Mg mass is relatively small, \about
$0.005~M_{\odot}$, which is, however, sufficient to form a strong
line, and the same holds for Ca.

The value of the $^{56}$Ni mass derived from these nebular models is
slightly larger than that derived from fitting the peak of the light
curve.  This is required mainly to fit the peak of the emission lines
of [\ion{O}{1}] and \ion{Mg}{1}].  However, these peaks are distinctly
narrower than the rest of the lines, as was the case for SN~1998bw.
This may indicate the presence of an inner concentration of matter (O,
Mg), a possible signature of an asymmetric explosion \citep{Maeda03}.


\begin{figure}
\ssp
\begin{center}
\scalebox{0.7}{
\plotone{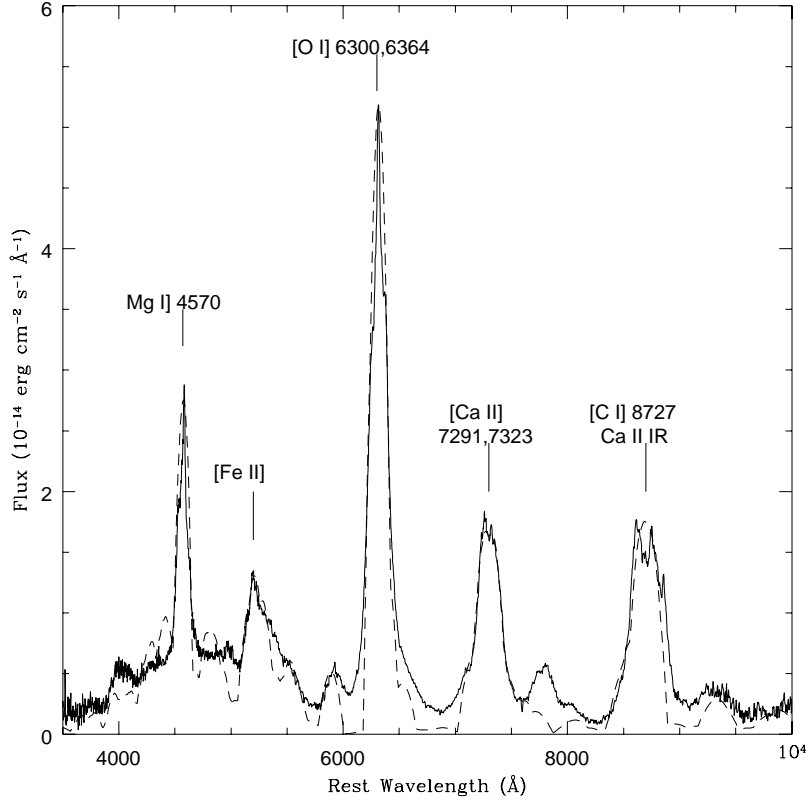}
}
\end{center}
\caption{The spectrum of SN2002ap from 2002 July 11.5 is compared to a
synthetic spectrum computed with a non-LTE model.  The model assumes
that the ejecta form a uniform sphere; it includes the effect of
heating via deposition of gamma rays and positrons, and cooling via
nebular line emission.  The model ejecta have an outer velocity of
5500\kms.  The mass of $^{56}$Ni used in the model ejecta is $0.09
M_{\sun}$, while the total mass within the outer velocity is $1.5
M_{\sun}$. This is somewhat larger than the value derived from the
explosion model used to compute synthetic light curves and spectra for
SN~2002ap at early times.}\label{f:latefit}
\end{figure}

Figure~\ref{f:latefit} shows a comparison of the nebular spectrum of
SN~2002ap observed on 2002 July 11, and a synthetic spectrum computed
with a NLTE, 1-zone code. The fiducial epoch of the spectrum is 163
days after explosion.

\section{DISCUSSION AND CONCLUSIONS}\label{s:disc}
The strong O and Mg emission, along with the lack of H and He, suggest
that SN~2002ap is a core-collapse event, where the hydrogen, helium,
and perhaps most of the carbon/oxygen layers are stripped.  The high
velocities in the early spectra further support this hypothesis since
more massive outer layers generally decelerate the ejected material
more than less massive outer layers.  In addition, stripping the outer
layers decreases the amount of ejecta, which in turn causes the
kinetic energy per unit of ejecta mass to increase.

The photometric and spectroscopic data suggest that SNe~2002ap,
1998bw, 1997ef, and 1994I are all related events.  However, SN~1994I
is distinctly different photometrically and spectroscopically from the
other three SNe.  The results of modeling of the spectra of these SNe
are presented in Table~\ref{t:params}.  These models suggest that
SNe~2002ap, 1998bw, and 1997ef may be unusually energetic SNe~Ic.
Because of their higher $^{56}$Ni mass, the gamma-ray trapping for SNe
1998bw and 1997ef was greater, resulting in broad light curves, while
the light curves of the normal SN~Ic~1994I are narrower and fade more
rapidly.  However, the $^{56}$Ni mass of SN~2002ap is not
significantly higher than that of SN~1994I, yet SN~2002ap has much
broader light curves.  In addition, SN~1999ex has broad light curves
without showing the high-velocity ejecta characteristic of SNe~1998bw
and 2002ap (for a discussion of fast and slow-declining SNe~Ib/c, see
\citealt{Clocchiatti97}).  SNe~2002ap and 1994I produced a similar
$^{56}$Ni mass, despite the slightly larger energy of the former.
Therefore, $^{56}$Ni mass is not a sufficient indicator of a peculiar
SN~Ic event similar to SN~1998bw.


\begin{deluxetable}{ccccccc}
\tabletypesize{\footnotesize}
\tablewidth{0pt}
\tablecaption{Model Parameters for SNe~2002ap, 1998bw, 1997ef, and 1994I\label{t:params}}
\tablehead{
\colhead{} &
\colhead{} &
\colhead{} &
\colhead{Kinetic Energy} &
\colhead{} \\
\colhead{Supernova} &
\colhead{$^{56}$Ni ($M_{\sun}$)} &
\colhead{Ejecta Mass ($M_{\sun}$)} &
\colhead{($10^{51}$ ergs)} &
\colhead{Reference}}

\startdata
SN~2002ap & \about 0.09 & \about 1.5  & $4 - 10$  & \citet{Mazzali02} \\
SN~1998bw & \about 0.7  & \about 10.9 & $20 - 50$ & \citet{Iwamoto98} \\
SN~1997ef & \about 0.13 & \about 9.6  & 17.5      & \citet{Mazzali00} \\
SN~1994I  & \about 0.07 & \about 0.88 & 1         & \citet{Nomoto94}
\enddata

\tablecomments{As noted by \citet{Hoflich99}, an asymmetric explosion
can reduce the energetics of these events.}

\end{deluxetable}

The lack of hydrogen and helium, along with the strong [\ion{O}{1}]
and \ion{Mg}{1}] lines in the nebular spectrum of SN~2002ap, suggest
that the progenitor's outermost layers were stripped.

The stripped nature of the progenitor of SN~2002ap can be further
investigated by examining the levels of magnesium and oxygen in its
nebular spectra.  A more heavily stripped star should reveal more of
the oxygen/magnesium/neon shell, and therefore less of the
carbon/oxygen shell, resulting in more magnesium (relative to oxygen)
present in the nebular spectrum.  The large Mg/O ratio of SN~2002ap
relative to other SNe~Ic indicates that we are seeing deep into the
core of the progenitor of SN~2002ap, suggesting that the progenitor's
carbon/oxygen layer was more stripped than the progenitors of most
SNe~Ic including SN~1998bw.  However, since no strong neon transitions
are seen in the nebular spectrum, it is inconclusive whether the
carbon/oxygen layer of the progenitor was completely stripped,
revealing the oxygen/neon/magnesium layer.  Modeling suggests that no
realistic star can contain enough neon to show a strong, unblended
neon line in the nebular phase.  In addition, mixing and excitation
effects may play a significant role in nebular [\ion{O}{1}] and
\ion{Mg}{1}] lines.  Perhaps if there were a strong unblended optical
carbon line, one could see \ion{Mg}{1}] increase and the carbon line
decrease with increased stripping.

It has been suggested that SN~1985F was of type SN~Ib
\citep{Matheson00}.  However, there are only late-time spectra of
SN~1985F, starting \about 160 d past $B$~maximum.  The similarity of
the SN~1985F spectrum in Figure~\ref{f:late_compare} to that of
SN~2002ap (specifically the [\ion{O}{1}]/[\ion{Ca}{2}] ratio), coupled
with the very broad $B$-band peak, suggests that SN~1985F was not a
SN~Ib and was probably a high-velocity SN~Ic.

Currently, there is a debate over how SNe~1998bw, 1997ef, 2002ap, and
their spectrally similar cousins should be classified (see
\citealt{Gal-Yam02a} for a discussion).  Many researchers have called
SN~2002ap a ``hypernova.''  However, the explosion energy of SN~2002ap
was probably only slightly higher than that of a normal SN~Ic, unlike
the case for the more powerful SN~1998bw.  Also, SN~2002ap showed no
evidence of relativistic ejecta at radio wavelengths \citep{Berger02}.
Moreover, the ``hypernova'' classification distracts casual readers of
the literature from the physical process involved in an event like
SN~2002ap.  Since ``hypernovae'' are probably the result of the
core-collapse of a massive star, we feel these events should still be
classified as supernovae.

The second portion of the classification discussion centers around
calling these events Type~Ic, Ic-peculiar, or Id.  Since the spectra
of SN~2002ap-like events are quite different from prototypical SN~Ic
spectra (such as SN~1994I), these events should not be classified as
normal Type~Ic.  However, the spectra of these events are not
drastically different from those of normal SNe~Ic.

This degree of similarity and heterogeneity is seen with peculiar
SNe~Ia.  In particular, despite the fact that the premaximum spectrum
of SN~1991T contained no \ion{Si}{2} or \ion{Ca}{2}, which are
defining characteristics of a SN~Ia, SN~1991T is considered a SN~Ia
and was not given its own classification of ``Ix.''  Furthermore, the
spectrum of SN~1991bg contained strong \ion{Ti}{2} features, deviating
from a typical SN~Ia spectrum, but SN~1991bg is still considered to be
a SN~Ia (see Filippenko 1997 for a review of these objects).  To
distinguish an object similar to the prototype SN~1991T (or SN~1991bg)
from ``normal'' SNe~Ia, we classify it as a SN~1991T-like (or a
SN~1991bg-like) SN~Ia.

Since SNe~2002ap, 1998bw, and 1997ef all have spectra similar to that
of SN~1994I (a typical SN~Ic), with the presence of high-velocity
lines being the largest difference, we feel it is unnecessary to use
the ``SN~Id'' designation for these objects.  Instead, by continuing
the syntax established with peculiar SNe~Ia, we propose that a SN with
a spectrum similar to that of SN~1998bw be classified as a
SN~1998bw-like SN~Ic.  Under this naming scheme, the correct
classification for SN~2002ap is that of a ``SN~1998bw-like SN~Ic.''

\begin{acknowledgments}
We thank the Lick Observatory and Keck Observatory staffs for their
assistance; Aaron Barth, Saurabh Jha, and Ed Moran also helped with
some of the observations. The W.~M. Keck Observatory is operated as a
scientific partnership among the California Institute of Technology,
the University of California, and NASA; the observatory was made
possible by the generous financial support of the W.~M. Keck
Foundation.  The work of A.V.F.'s group at the University of
California, Berkeley, is supported by National Science Foundation
grant AST-9987438, as well as by the Sylvia and Jim Katzman
Foundation.  Additional funding is provided by NASA through grants
GO-9114, GO-9155 and GO-9428 from the Space Telescope Science
Institute, which is operated by the Association of Universities for
Research in Astronomy, Inc., under NASA contract NAS~5-26555.  KAIT
was made possible by generous donations from Sun Microsystems, Inc.,
the Hewlett-Packard Company, AutoScope Corporation, Lick Observatory,
the National Science Foundation, the University of California, and the
Katzman Foundation.  P.M.'s work has been supported in part by the
Grant-in-Aid for Scientific Research (14047206, 14540223, 15204010) of
the Ministry of Education, Science, Culture, Sports, and Technology in
Japan.

\end{acknowledgments}


\clearpage

\begin{thebibliography}{}

\bibitem[Barbon et~al.(1990)]{Barbon90} Barbon, R., Benetti, S., 
Rosino, L., Cappellaro, E., \& Turatto, M.\ 1990, \aap, 237, 79

\bibitem[Berger, Kulkarni, \& Chevalier(2002)]{Berger02} Berger, 
E., Kulkarni, S.~R., \& Chevalier, R.~A.\ 2002, \apjl, 577, L5

\bibitem[Bessell(1999)]{Bessell99}Bessell, M.~S.\ 1999, \pasp, 111,
1426

\bibitem[Chornock et~al.(2003)]{Chornock03}Chornock, R., Foley, R.~J.,
Filippenko, A.~V., Papenkova, M., \& Weisz, D.\ 2003, \iaucirc\ 8114

\bibitem[Clocchiatti \& Wheeler(1997)]{Clocchiatti97} Clocchiatti, 
A.,~\& Wheeler, J.~C.\ 1997, \apj, 491, 375

\bibitem[Clocchiatti et al.(2001)]{Clocchiatti01} Clocchiatti, A.,
et~al.\ 2001, \apj, 553, 886

\bibitem[Filippenko(1982)]{Filippenko82}Filippenko, A.~V.\ 1982,
\pasp, 94, 715

\bibitem[Filippenko(1989)]{Filippenko89} Filippenko, A.~V.\ 1989, 
\aj, 97, 726

\bibitem[Filippenko(1997)]{Filippenko97}Filippenko, A.~V.\ 1997,
\araa, 35, 309

\bibitem[Filippenko(2003)]{Filippenko03}Filippenko, A.~V.\ 2003, in
>From Twilight to Highlight: The Physics of Supernovae, ed. W.
Hillebrandt \& B. Leibundgut (Berlin: Springer-Verlag), 171

\bibitem[Filippenko \& Chornock(2002)]{Filippenko02} Filippenko, 
A.~V.,~\& Chornock, R.\ 2002, \iaucirc\ 7825

\bibitem[Filippenko et~al.(2001)]{Filippenko01}Filippenko, A.~V., Li, W.~D.,
Treffers, R.~R., \& Modjaz, M.\ 2001, in Small-Telescope Astronomy on
Global Scales, ed. W.~P. Chen, C. Lemme, \& B. Paczy\'{n}ski (San
Francisco: Astron. Soc. Pacific, Conf. Ser. Vol. 246), 121

\bibitem[Filippenko \& Sargent(1986)]{Filippenko86} Filippenko, 
A.~V., \& Sargent, W.~L.~W.\ 1986, \aj, 91, 691

\bibitem[Filippenko et~al.(1995)]{Filippenko95} Filippenko,
A.~V., et~al.\ 1995, \apjl, 450, L11

\bibitem[Galama et~al.(1998)]{Galama98}Galama, T.~J., et~al.\ 1998,
Nature, 395, 670-672

\bibitem[Gal-Yam, Ofek, \& Shemmer(2002)]{Gal-Yam02a} Gal-Yam, A., 
Ofek, E.~O., \& Shemmer, O.\ 2002, \mnras, 332, L73

\bibitem[Gal-Yam, Shemmer, \& Dann(2002)]{Gal-Yam02b} Gal-Yam, A., 
Shemmer, O., \& Dann, J.\ 2002, \iaucirc\ 7811

\bibitem[Garnavich et~al.(2003)]{Garnavich03a}Garnavich, P., Matheson,
T., Olszewski, E.~W., Harding, P., \& Stanek, K.~Z.\ 2003, \iaucirc\
8114

\bibitem[Garnavich (2003)]{Garnavich03b} Garnavich, P. 2003,
private communication

\bibitem[Henden(2002)]{Henden02}Henden, A. 2002, GCN Circ. 1242

\bibitem[H\"{o}flich(1991)]{Hoflich91} H\"{o}flich, P.\ 1991, \aap, 246,
481

\bibitem[H\"{o}flich et~al.(1999)]{Hoflich99} H\"{o}flich, P.,
Wheeler, J.~C., \& Wang, L.\ 1999, \apj, 521, 179

\bibitem[Iwamoto et~al.(1998)]{Iwamoto98} Iwamoto, K., et~al.\ 1998,
\nat, 395, 672

\bibitem[Iwamoto et~al.(2000)]{Iwamoto00} Iwamoto, K., et~al.\ 2000,
\apj, 534, 660

\bibitem[Kawabata et~al.(2002)]{Kawabata02} Kawabata, K.~S., et~al.\
2002, \apjl, 580, L39

\bibitem[Kinugasa et~al.(2002)]{Kinugasa02} Kinugasa, K., 
Kawakita, H., Ayani, K., Kawabata, T., \& Yamaoka, H.\ 2002, \iaucirc\ 
7811

\bibitem[Klose, Guenther, \& Woitas(2002)]{Klose02} Klose, S.,
Guenther, E., \& Woitas, J.\ 2002, GRB Circular Network, 1248

\bibitem[Leonard et~al.(2002)]{Leonard02}
Leonard, D.~C., Filippenko, A.~V., Chornock, R., \& Foley, R.~J.\
2002, \pasp, 114, 1333

\bibitem[Li et~al.(2001)]{Li01}Li, W.~D., et~al.\ 2001, \pasp, 113,
1178

\bibitem[Maeda \& Nomoto(2003)]{Maeda03} Maeda, K., \& Nomoto, 
K.\ 2003, \apj, submitted (astro-ph/0304172)

\bibitem[Maeda et~al.(2003)]{Maeda03b} Maeda, K., Mazzali, P.~A.,
Deng, J., Nomoto, K., Yoshii, Y., Tomita, H., \& Kobayashi, Y. \ 
2003, \apj, in press (astro-ph/0305182)

\bibitem[Matheson(2000)]{Matheson00} Matheson, T.\ 2000,
Ph.D.~thesis, University of California, Berkeley

\bibitem[Matheson et~al.(2000a)]{Matheson00a} Matheson, T.,
Filippenko, A.~V., Chornock, R., Leonard, D.~C., \& Li, W.\ 2000a, \aj,
119, 2303

\bibitem[Matheson et~al.(2000b)]{Matheson00b} Matheson, T.,
Filippenko, A.~V., Ho, L.~C., Barth, A.~J., \& Leonard, D.~C.\ 2000b,
\aj, 120, 1499

\bibitem[Matheson et~al.(2001)]{Matheson01}Matheson, T., Filippenko,
A.~V., Li, W., \& Leonard, D.~C.\ 2001, \aj, 121, 1648

\bibitem[Matheson et~al.(2000c)]{Matheson00c} Matheson, T., et~al.\ 
2000c, \aj, 120, 1487

\bibitem[Mazzali et~al.(2000)]{Mazzali00}Mazzali, P.~A., Iwamoto, K.,
\& Nomoto, K.\ 2000, \apj, 545, 407

\bibitem[Mazzali et~al.(2001)]{Mazzali01} Mazzali, P.~A., Nomoto, K.,
Patat, F., \& Maeda, K.\ 2001, \apj, 559, 1047

\bibitem[Mazzali et~al.(2002)]{Mazzali02}Mazzali, P.~A., et~al.\ 2002,
\apjl, 572, L61

\bibitem[Mazzali et~al.(2003)]{Mazzali03}Mazzali, P.~A., Nomoto, K., Deng, J., 
Maeda, K., \& Qiu, Y. \ 2003, in From Twilight to Highlight: The 
Physics of Supernovae, ed. W. Hillebrandt \& B. Leibundgut (Berlin: 
Springer-Verlag), 246

\bibitem[Meikle et~al.(2002)]{Meikle02} Meikle, P., Lucy, L., Smartt,
S., Leibundgut, B., Lundqvist, P., \& Ostensen, R.\ 2002, \iaucirc\
7811

\bibitem[Miller \& Stone(1993)]{Miller93}Miller, J. S., \& Stone,
R.~P.~S.\ 1993, Lick Obs. Tech. Rep.\ No. 66 (Santa Cruz, CA: Lick Obs.)

\bibitem[Modjaz et~al.(2000)]{Modjaz00}Modjaz, M., Li, W.~D.,
Filippenko, A.~V., King, J.~Y., Leonard, D.~C., Matheson, T.,
Treffers, R.~R., \& Riess, A.~G.\ 2000, \pasp, 113, 308

\bibitem[Nakano et~al.(2002)]{Nakano02}Nakano, S., Kushida, R.,
Kushida, Y., \& Li, W.\ 2002, \iaucirc\ 7810

\bibitem[Nomoto et~al.(1994)]{Nomoto94} Nomoto, K., Yamaoka, H., Pols,
O.~R., van den Heuvel, E.~P.~J., Iwamoto, K., Kumagai, S., \&
Shigeyama, T.\ 1994, \nat, 371, 227

\bibitem[O'Donnell(1994)]{ODonnell94} O'Donnell, J.~E.\ 1994, 
\apj, 422, 158 

\bibitem[Oke et~al.(1995)]{Oke95} Oke, J.~B., et~al.\ 1995, \pasp,
107, 375

\bibitem[Pandey et~al.(2003)]{Pandey03}Pandey, S.~B., Anupama, G.~C.,
Sagar, R., Bhattacharya, D., Sahu, D.~K., \& Pandey, J.~C.\ 2003,
\mnras, submitted (astro-ph/0209507)

\bibitem[Patat et~al.(1993)]{Patat93}Patat, F., Barbon, R.,
Cappellaro, E., \& Turatto, M.\ 1993, \aaps, 98, 443

\bibitem[Richmond (2003)]{Richmond03}Richmond, M.~W. 2003,
private communication

\bibitem[Richmond et~al.(1996)]{Richmond96}Richmond, M.~W., Treffers,
R.~R., Filippenko, A.~V., \& Paik, Y.\ 1996, \aj, 112, 732

\bibitem[Schlegel, Finkbeiner, \& Davis(1998)]{Schlegel98} Schlegel,
D.~J., Finkbeiner, D.~P., \& Davis, M.\ 1998, \apj, 500, 525  

\bibitem[Sharina, Karachentsev, \& Tikhonov(1996)]{Sharina96} 
Sharina, M.~E., Karachentsev, I.~D., \& Tikhonov, N.~A.\ 1996, \aaps, 119, 
499 

\bibitem[Sheinis et~al.(2002)]{Sheinis02} Sheinis, A.~I., Bolte,  M.,
Epps, H.~W., Kibrick, R.~I., Miller, J.~S., Radovan, M.~V., Bigelow,
B.~C., \& Sutin, B.~M.\ 2002, \pasp, 114, 851

\bibitem[Sohn \& Davidge(1996)]{Sohn96} Sohn, Y., \& Davidge, 
T.~J.\ 1996, \aj, 111, 2280 

\bibitem[Sollerman et~al.(2000)]{Sollerman00} Sollerman, J., Kozma, 
C., Fransson, C., Leibundgut, B., Lundqvist, P., Ryde, F., \& Woudt, P.\ 
2000, \apjl, 537, L127

\bibitem[Sollerman et~al.(2002)]{Sollerman02} Sollerman, J., et~al.\ 
2002, \aap, 386, 944 

\bibitem[Stanek et al.(2003)]{Stanek03} Stanek, K.~Z., et~al.\ 
2003, \apjl, in press (astro-ph/0304173)

\bibitem[Stathakis et~al.(2000)]{Stathakis00} Stathakis, R.~A.,
et~al.\ 2000, \mnras, 314, 807

\bibitem[Stetson(1987)]{Stetson87} Stetson, P.~B.\ 1987, \pasp, 99, 191

\bibitem[Stritzinger et al.(2002)]{Stritzinger02} Stritzinger, M.,
et~al.\ 2002, \aj, 124, 2100

\bibitem[Sutaria et~al.(2003)]{Sutaria03} 
Sutaria, F.~K., Chandra, P., Bhatnagar, S., \& Ray, A.\ 2003, \aap,
397, 1011

\bibitem[Thomas et~al.(2003)]{Thomas03} Thomas, R.~C., Branch, 
D., Baron, E., Nomoto, K., Li, W., \& Filippenko, A.~V.\ 2003, \apj,
submitted (astro-ph/0302260)

\bibitem[Tsvetkov(1986)]{Tsvetkov86}Tsvetkov, D.~Y.\ 1986, SvAL, 12,
328

\bibitem[Wang et~al.(2003)]{Wang03}Wang, L., Baade, D., H\"{o}flich, P.,
Wheeler, J. C., Fransson, C., \& Lundqvist, P.\ 2003, \apj, submitted
(astro-ph/0206386) 


\bibitem[Yoshii et~al.(2003)]{Yoshii03}Yoshii, Y., et~al.\ 2003, \apj,
accepted (astro-ph/0304010)

\end{thebibliography}
\end{document}